\documentclass[journal=jctcce,manuscript=article]{achemso}
\usepackage{amssymb}
\usepackage{amsmath}
\usepackage{wasysym}
\usepackage{graphicx}
\usepackage{bm}
\usepackage{booktabs}
\usepackage{multirow}
\usepackage{color}

\makeatletter
\g@addto@macro\bfseries{\boldmath}
\makeatother

\newcommand{\GoWo}{\text{G}_0\text{W}_0}

\newcommand{\frg}{$\text{FRG }$}

\newcommand{\figref}[1]{\mbox{Figure~\ref{#1}}}
\newcommand{\tabref}[1]{\mbox{Table~\ref{#1}}}

\newcommand{\np}{n'}
\newcommand{\nun}{\underline{n}}

\newcommand{\idn}{_{n}}

\newcommand{\hamh}{H_{\text{H}}}          

\newcommand{\self}{\Sigma}             

\newcommand{\eqp}{\varepsilon}        

\newcommand{\bracetop}[3]{\left<#1\right\vert#2\left\vert#3\right>}
\newcommand{\ci}{\mathfrak{i}}
\newcommand{\br}{\mathbf{r}}
\newcommand{\be}{\begin{equation}}

\newcommand{\bea}{\begin{eqnarray}}
\newcommand{\eea}{\end{eqnarray}}
\newcommand{\mA}{\mathfrak{A}}

\newcommand{\IP}{I_{\text{P}}}
\newcommand{\EA}{E_{\text{A}}}

\title{Quasi-Particle Self-Consistent $GW$ for Molecules.}
\author{F. Kaplan}
\affiliation{Institute of Nanotechnology, Karlsruhe Institute of Technology, Campus North, D-76344 Karlsruhe, Germany}
\author{M. E. Harding}
\affiliation{Institute of Nanotechnology, Karlsruhe Institute of Technology, Campus North, D-76344 Karlsruhe, Germany}
\author{C. Seiler}
\affiliation{Institute of Theoretical Physics, University of Regensburg, D-93040 Regensburg, Germany}
\author{F. Weigend}
\affiliation{Institute of Nanotechnology, Karlsruhe Institute of Technology, Campus North, D-76344 Karlsruhe, Germany}
\altaffiliation{Institute of Physical Chemistry, Karlsruhe Institute of Technology, Campus South, D-76021 Karlsruhe, Germany}
\author{F. Evers}
\affiliation{Institute of Theoretical Physics, University of Regensburg, D-93040 Regensburg, Germany}
\author{M. J. {van Setten}}
\email{mjvansetten@gmail.com}
\affiliation{Nanoscopic Physics, Institute of Condensed Matter and Nanosciences, Universit\'{e} Catholique de Louvain, 1348 Louvain-la-Neuve, Belgium}

\date{\today}

\begin{document}

\begin{abstract}
We present the formalism and implementation of quasi-particle self-consistent $GW$ (qs$GW$) and eigenvalue only quasi-particle self-consistent $GW$ (ev$GW$) adapted to standard quantum chemistry packages. Our implementation is benchmarked against high-level quantum chemistry computations (coupled-cluster theory) and experimental results using a representative set of molecules. Furthermore, we compare the qs$GW$ approach for five molecules relevant for organic photovoltaics to self-consistent $GW$ results (sc$GW$) and analyze the effects of the self-consistency on the ground state density by comparing calculated dipole moments to their experimental values. We show that qs$GW$ makes a significant improvement over conventional $G_0W_0$ and that partially self-consistent flavors (in particular ev$GW$) can be excellent alternatives.
\end{abstract}

\maketitle

\section{Introduction}

The sufficiently accurate prediction of photo-ionization processes is still a serious computational challenge. The main workhorse for medium and large sized systems is the Kohn-Sham density functional theory (KS-DFT).\cite{dft, kohn65} As is very well known, using KS-DFT eigenvalues (especially using semi local functionals) for computational spectroscopy has, however, various fundamental and practical limitations. Moreover $\Delta$SCF ionisation energies, which are more accurate, are in general only applicable to obtain the first ionisation energies.\cite{chonggritsenkobaerends02,gritsenkobaerends09,baerends13} An approach promising better accuracy for the calculation of single particle excitations is the $GW$-method. Its central object is the Green's function $G$.\cite{hedin65,hedin65a, hedin71, hedin99} It is calculated by solving the Dyson equation that relates the full (interacting) Green's function to a know non-interacting reference one, $G_0$, via a self-energy $\Sigma$. Depending on the starting point, $G_0$, $\Sigma$ adds or corrects exchange and (dynamical) correlation. In $GW$-theory $\Sigma$ is approximated evaluating the Fock diagram, however employing a screened interaction $W$. Since the Green's function, $G$, exhibits (complex) poles that describe the (charged) excitation energies (and their lifetimes),\cite{abrikosov63,fetter71} $GW$ theory represents a simple and transparent framework for the investigation of ionization processes. 

Since $GW$ comes with a substantial computational effort, a great variety of simplified flavors of $GW$ are in use, the most common being so called $G_0W_0$ approximation.\cite{hedin99} It treats the self-energy $\Sigma$ as a first order (i.e. nonselfconsistent) perturbation acting on a KS or Hartree Fock (HF) reference system. It produces sizeable corrections, in particular to the energies of the frontier orbitals, at moderate computational costs; thus medium sized molecules can be treated efficiently. These advantages are reflected in an increasing number of applications in quantum chemistry.\cite{zaider96prb, ishii01prb63, loui01, PhysRevB.65.245109, PhysRevLett.90.076401, hubner04pla, tiago06, puerto06, puerto08, ramos08, noguchi08, bruneval09, tiago09, thygesen10-mol, olevano11prb, xiaofeng11, PhysRevB.84.075144, sanchezportal11jcp, ke11prb, rohlfing12jctc, neaton12mo, nguyen12, aimsgw,   PhysRevB.85.245125,   bruneval/atom,  pham13, bruneval13jctc,  PhysRevB.83.115123, vansetten13, faber13032014, faber2013jcp, baumeier2014, ren15, migani15, abinitsternh, kuhn15, vansetten15gw100, kaplan15gwso}

However, the lack of self-consistency in $G_0W_0$ implicates undesirable shortcomings: Both, $G$ and the polarization $P$, that enter the calculation are imported without change from the underlying reference calculation (KS or HF). As a consequence, $G_0W_0$ depends on the choice of the reference systems, e.g., the exchange-correlation (XC) functional of KS-DFT calculation. Furthermore, there is no update in the spatial shape of the orbitals, so the ground state density of $G_0W_0$ reproduces the one of the reference theory. 

These limitations are overcome by imposing self-consistency. However, due to the fact that fully self-consistent $GW$ (sc$GW$) is computationally expensive,\cite{wang15, caruso12prb, thygesen10, thygesen10-mol, aimsgw, Sakuma/Miyake/Aryasetiawan:2009,  Stan2009, Holm1998a, caruso13prb, gatti07, koval14} it is highly desirable to explore the potential of partially self-consistent schemes. In this work we follow a procedure towards partial self-consistency that is numerically still tractable and gives promising results; the quasi-particle (QP) self-consistent GW (qs$GW$). For solids there is already encouraging experience with this approach,\cite{vanschilfgaarde06prl, vanschilfgaarde04prl, vanschilfgaarde06, kotani07, shishkin:235102, shishkin:246403, bruneval06prb, bruneval06prl} and first applications to atoms\cite{bruneval/atom} and small molecules\cite{koval14} also seem promising. Here it is adapted for use in a general quantum chemistry code. Our results indicate that the starting point dependence, observed at the $G_0W_0$ level, which can easily exceed 1.0~eV,\cite{Holm/vonBarth:2004,rinke05njp,Fuchs/etal:2007,Koerzdoerfer/Marom:2012,caruso12prb,Marom/etal:2012,bruneval13jctc,Atalla/etal:2013,caruso12prb,thygesen10-mol,vansetten13,marom2012,koval14,lischner/etal:2014} is completely removed in qs$GW$. 

Our calculations suggest that the effects of orbital updates can be often neglected in comparison to the correction coming from the shifts in the 
 pole positions 
of the Green's function. Therefore, we also implement and investigate a simplified version of qs$GW$ in which the orbitals are kept fixed at the reference (DFT of HF) result and only the QP-energies are updated (eigenvalue only $GW$, ev$GW$). Also this, and similar schemes, have been successful already in implementations for solids\cite{vaspgw3, vaspgw4, abinit} and first applications in molecular geometries.\cite{faber13032014, olevano11prb, kaplan15gwso} It is an approximation especially beneficial for larger systems since it is computationally less demanding than sc$GW$ (and qs$GW$). An important aspect of our work is that we present a systematic comparison between qs$GW$ and ev$GW$ providing a validation of ev$GW$. In particular we find that in ev$GW$ the starting point dependence is strongly reduced as compared to $G_0W_0$.

An important aspect of testing a new methodology is comparison with other approaches. A generally popular reference are  experimental values. This commonly adopted practice can however be misleading in the case of ionization energies. One difficulty is that the experimental ionization energies are often adiabatic whereas those calculated within the GW scheme are always vertical. In addition intrinsic effects originating from zero-point vibrations and relativistic effects (beyond those that enter via the reference calculation) are usually not included, also not in our current approach. (We mention that at the $G_0W_0$ level a two component extension has been implemented recently to account for spin-orbit effect for closed shell molecules.\cite{kuhn15}) Therefore a comparison to more accurate theoretical results would be more reliable. In this work, we make use of the possibility to compare our results to coupled-cluster singles and doubles augmented by a perturbative treatment of triple excitations (CCSD(T))\cite{purvis82, raghavachari89} employing the same atomic structure and basis set.\cite{krause15} Doing so rules out experimental uncertainties, temperature and zero-point renormalization effects etc., but also basis set errors. 

Our paper is organized as follows: In section~\ref{sec:qsgw} we present the formalism that is used within our implementation of qs$GW$. Section~\ref{sec:implementation} explains the details of the implementation. The third section validates the qs$GW$ method. First, our implementation is tested for internal consistency (Sec.~\ref{sect-internal}), i.e. we test the convergence behavior with the basis set and the number of iteration cycles. Second, the qs$GW$ method is assessed using first ionization potentials (IP) of a set of 29 representative molecules (section~\ref{sect-1stIPqs}).\cite{vansetten13, caruso12prb} Third, we compare higher IPs with experimental data (section~\ref{sect-higherIPqs}) and with results using the sc$GW$ implementation within the FHI-AIMS software package\cite{caruso12prb} (section~\ref{sect-qsvssc}). Furthermore, we evaluate the ground state densities and the dipole moments in section~\ref{sect-density}. In section~\ref{sect-g0w0comparison} we compare the results of qs$GW$ to those of $G_0W_0$. 

In section~\ref{sect-partialsc} we investigate two different approaches of partial self-consistency. We first introduce and benchmark our implementation of the ev$GW$ method (self-consistency in the poles positions of $G$ and the response function) versus qs$GW$ and CCSD(T). In section~\ref{sect-gevw0} we test the $G_{ev}W_0$\cite{kaplan15gwso} approach (self-consistent only in the poles positions of $G$), versus qs$GW$. Finally section~\ref{sect-compperf} contains an analysis of the computational performance and scaling behavior of the presented methods.

\section{Method} \label{sec:method}

\subsection{qs$GW$ Approximation} \label{sec:qsgw} 

We are aiming at an approximate self-consistent solution of the Dyson equation\begin{equation} \label{eq:scG}
G(E) = \left( E - H[G] -\self[G] \right)^{-1}
\end{equation} 
Here, $H[G]$ denotes the Hartree contribution incorporating the external potential (respectively the ions), the kinetic contribution and the Hartree potential. The latter depends on the charge density and hence is considered as a functional of $G$. 

In the $GW$-approximation\cite{hedin99} the self-energy is given in terms of the causal Green's function:
\begin{equation}\label{eq:selfFreq}
\self(\br, \br', \omega) = \frac{\ci}{2 \pi} \int  \text{d}\omega' G(\br, \br', \omega + \omega')W(\br,\br',\omega') \text{e}^{\ci \eta \omega'} 
\end{equation}
and the screened Coulomb interaction $W$. $\eta$ denotes a positive infinitesimal. $W$ is obtained from the following equation
\begin{equation}\label{eq:screenedInteraction}
W(\br, \br', \omega) =  v(\br, \br') + \int \text{d}\br'' \int \text{d}\br''' v(\br, \br'') P(\br'', \br''', \omega)W(\br''', \br',\omega) 
\end{equation}
with $v(\br,\br')=e^2\left|\br-\br'\right|^{-1}$ denoting the bare Coulomb kernel. Ignoring vertex corrections (consistent with the construction of $\self$) the polarization $P$ is calculated within the random phase approximation (RPA)
\begin{equation}\label{eq:polarization}
P_{\text{RPA}}(\br, \br', \omega) = -\frac{\ci}{2 \pi}  \int \text{d}\omega' G(\br,\br',\omega + \omega')G(\br,\br',\omega') \text{e}^{\ci \eta \omega'} 
\end{equation}

\subsubsection{Quasi-Particle Equation and Quasistatic Approximation}

A solution of equation \ref{eq:scG} is constructed by introducing a quasi-spectral representation; 
\begin{align} \label{eq:qsGF}
G(\br,\br',z) = \sum_{n} \frac{\psi_{r,n}(\br,z) \bar{\psi}_{l,n}(\br',z)}{z - \eqp_{n}(z) + \ci \eta \text{sgn}(\eqp_{n}(z) - \mu)} 
\end{align}
The right and left eigenvectors, $\psi_{r,n}(\br,z)$, $\psi_{l,n}(\br,z)$ and eigenvalues $\eqp_{n}(z)$ represent quasi-particle/hole (QP/QH) states and energies, $\mu$ being the chemical potential. The bar in $\bar{\psi}_{l,n}(\br',z)$ denotes the complex conjugate. The (complex) poles $z$ of the Green's function \ref{eq:qsGF} follow from the pole condition
\begin{equation}
z - \eqp_{n}(z) = 0 
\end{equation}
The QP-orbitals as well as the energies are found as solutions of the QP-equations
\begin{align} 
\left (\eqp\idn(z) - \hamh[G]-\self[G] \right ) \cdot \psi_{r,n}(\br, z) = 0 \\
\psi_{l,n}(\br, z) \cdot \left (\eqp\idn(z) - \hamh[G] -\self[G] \right )  = 0 
\end{align}

Within qs$GW$ the self-energy \ref{eq:selfFreq} is approximated by an energy-independent, Hermitian matrix. In the literature different variants have been proposed for such approximate self-energies.\cite{vanschilfgaarde04prl,vanschilfgaarde06, sakuma09} Here we follow Faleev {\em et al.}\cite{vanschilfgaarde04prl} 
\begin{equation}\label{eq:effSelf}
 \tilde{\Sigma}_{nn'} = \frac{1}{2} \left( \self_{n \np}(\eqp_{n}) +  \self_{n \np}(\eqp_{\np})\right) 
\end{equation}

This approximation takes into account the quasi-particle part of the Green's function and neglects life-time effects. One of its merits is that it implies a consistent treatment of the renormalization factor $Z$ within the calculation of many observables.\cite{vanschilfgaarde04prl} 

Removing the energy dependence from the self-energy has important computational benefits. First, the energy integration in eq~\ref{eq:selfFreq} can be performed analytically. Second, the approximate self-energy takes an Hermitian form hence left and right eigenvectors coincide
\begin{align} \label{eq:QPE}
\left (\eqp\idn - \hamh[G] -\tilde{\Sigma}[G] \right ) \cdot \psi_{n}(\br) = 0 
\end{align}
and the poles are real, reflecting an effective single particle theory. Consequently, the Green's function of qs$GW$ takes the form
\begin{align}\label{qsgwgf}
G = \sum_n \frac{\psi\idn(\br) \bar{\psi}\idn(\br')}{E - \eqp_{n} + \ci \eta \text{sgn}(\eqp_n - \eta)}
\end{align}

\subsubsection{Kohn-Sham Initialization}

An iterative procedure of solving eq~\ref{eq:QPE} and \ref{qsgwgf} self-consistently is typically initialized with a KS Green's function constructed from the KS orbitals $\psi^{(0)}\idn(\br)$ and energies $\eqp^{(0)}_n$:
\begin{align}\label{init}
G^{(0)}(\br,\br',E) = \sum_{n} \frac{\psi^{(0)}\idn(\br) \bar{\psi}^{(0)}\idn(\br')}{E - \eqp^{(0)}_{n} + \ci \eta \text{sgn}(\eqp^{(0)}_n - \eta)} 
\end{align}
 
\subsection{Implementation}\label{sec:implementation}

qs$GW$ has been implemented within a local version of the TURBOMOLE package, building on the routines for the calculation of the $G_0W_0$ self-energy.\cite{vansetten13} 

The solution of eq~\ref{eq:QPE} is organized in an iterative scheme starting from the KS-initialization eq~\ref{init}. The QP-orbitals of the $(i+1)$th iteration $\psi^{(i+1)}_{n}(\br) $ are expressed in the reference orbitals of the previous iteration:
\begin{equation} \label{eq:ks2qp}
\psi^{(i+1)}_{n}(\br) = \sum_{\nun}  \mA^{(i+1)}_{n \underline{n}} \psi^{(i)}_{\nun}(\br)
\end{equation}
In the reference basis $\psi^{(i)}_{\nun}(\br)$ eq~\ref{eq:QPE} takes the form of an eigenvalue problem
\begin{multline}\label{eq:qpe2}
\sum_{\nun} \mA^{(i+1)}_{n' \nun} \bigg [ \int \text{d}\br' \int \text{d}\br \phantom{.}\psi^{(i)}_{n}(\br) \big( \hamh[G^{(i)}] \delta(\br-\br') + \tilde{\Sigma}(\br, \br') \big) \psi^{(i)}_{\nun}(\br')  \bigg ] = \eqp^{(i+1)}_{n'}\mA^{(i+1)}_{n' n}
\end{multline}
The diagonalization of eq~\ref{eq:qpe2} yields updates in $\eqp^{(i+1)}_{n'}$ and $\mA^{(i+1)}_{n' n}$. With the latter new orbitals $\psi^{(i+1)}_{n}(\br)$ are constructed via \ref{eq:ks2qp}. These are orthonormal by construction due to the hermiticity of the operators in eq~\ref{eq:qpe2}.

Since hermiticity gives the present scheme a form similar to an effective single particle problem we can take advantage of established (DFT) routines to calculate the Hartree Hamiltonian $\hamh[n^{(i+1)}]$, $n^{(i+1)} = \sum_{i=1}^N\left|\psi^{(i+1)}_{n}(\br)\right|^2$ employing an updated density.

For the computation of the matrix elements of the self-energy 
\begin{equation}
\self_{n \nun}(E) = \bracetop{n}{\Sigma_{\text{x}}}{\nun} + \bracetop{n}{\Sigma_{\text{c}}(E)}{\nun}
\end{equation}
we recall the expressions already derived before in the context of the $G_0W_0$ implementation.\cite{vansetten13} The expression for the real part of a matrix element of the correlation part $\Sigma_{\text{c}}$ of the self-energy reads;
\begin{equation}
\begin{aligned}
 &\text{Re} \big( \bracetop{n}{\Sigma_{\text{c}}(E)}{n'} \big)  = \\ \sum_{m} \bigg [  & \sum_{i}^{\text{occ}} (i n | \rho_{m}) (\rho_{m} | n' i ) \times \frac{E -\eqp_{i} + \Omega_m}{(E -\eqp_{i} + \Omega_m)^2 + \bar{\eta}^{2}} \\  + & \sum_{a}^{\text{unocc}}  (a n | \rho_{m}) (\rho_{m} | n' a ) \times \frac{E -\eqp_{a} - \Omega_m}{(E -\eqp_{a} - \Omega_m)^2 + \bar{\eta}^{2}} \bigg ] 
\end{aligned} \label{eq:selfC_matelem}
\end{equation}
where $\rho_m$ and $\Omega_m$ denote the two particle excitation densities and energies and $\bar{\eta}$ is a positive infinitesimal. To express the exchange part we employ the common notation of the Coulomb integrals
\begin{equation}
(p q| r s) = \int \text{d}\br \int \text{d} \br' p(\br) q(\br) \frac{1}{| \br - \br' |} r(\br') s(\br')  \label{eq:cexint}
\end{equation}
$p(\br)$, $q(\br)$, etc. denoting single particle orbitals, e.g. of QP or KS type.

The unscreened exchange part of qs$GW$ is identical to the exchange contribution of HF theory,
\begin{equation}
\bracetop{n}{\Sigma_x}{{n'}} = \sum_{\nun}^{\text{occ}} (n \nun| \nun n')  \label{eq:exchange}
\end{equation}
where the sum is over all occupied (QP-)orbitals.

Employing the quasi-spectral representation, eq~\ref{eq:qsGF}, the new self-energy is calculated from the updated 
pole positions $\eqp^{(i+1)}_{n'}$ and orbitals $\psi^{(i+1)}_{n}(\br)$. This uses the established routines from the $G_0W_0$ implementation, taking into account all changes, i.e. in both $G$ and $W$.

After the approximation eq~\ref{eq:effSelf} is applied the next iteration is started by solving again eq~\ref{eq:qpe2}. This procedure is continued until a self-consistent solution is achieved. 

\subsubsection{Convergence Criteria} 

In terminating the self-consistency cycle different convergence criteria are conceivable. An obvious choice would consider the change of the QP energies from one cycle to the next. However, motivated by earlier work,\cite{caruso12prb} we check for the norm of the differences of the Green's functions:
\begin{align}
 \Delta & = \frac{1}{N^{2}_{\text{Orbitals}}}\sum_{n, \nun} \left| G_{n \nun}(E=0) - G^{(i-1)}_{n \nun}(E=0) \right| \\
 & = \frac{1}{N^{2}_{\text{Orbitals}}}\sum_{n} \left| G_{n n}(E=0) - G^{(i-1)}_{n n}(E=0) \right|.
 \end{align}
The last step is valid when $G_{n,\nun}$ is diagonal, i.e. given in the orthonormal eigenstates of the QP-equation. Typically the iteration is stopped if $\Delta < 10^{-7}$ is achieved. This corresponds to orbital energies being converged to within 1~meV.

\subsubsection{Linear Mixing} 
\label{sect:linmix}
\begin{figure}[!htb]
   \includegraphics[width=0.95\columnwidth]{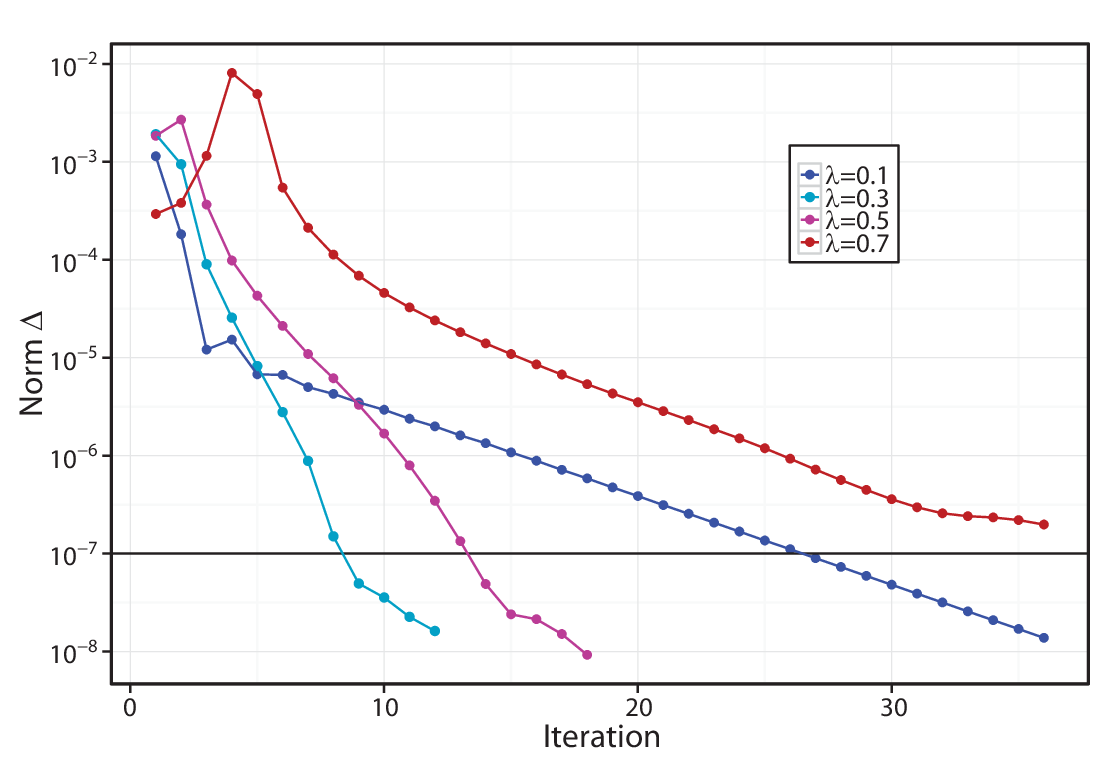} 
  \caption{ The flow of the convergence of the norm of the difference of the Green's functions from two iterations with the qs$GW$ iteration for the BF. The termination criterion is indicated by the solid line.}
  \label{fig:norm_bf}
\end{figure}

The iterative approach introduced above does not always converge into a fixed point solution. To improve stability and also rate of convergence we introduce a linear mixing scheme that mixes into the updated Green's function a contribution of the previous one to decrease the step width between two iterations in a similar manner as done for iterative procedures for the HF or DFT ground state. Therefore, strictly speaking we do not solve eq~\ref{eq:qpe2}, but 
\begin{equation}
\bigg [\eqp^{(i+1)}\idn - \Big( \lambda \big(\hamh[G^{(i)}] + \self[G^{(i)}]\big) + (1-\lambda)\eqp^{(i)}\idn \Big)  \bigg ]\psi^{(i+1)}\idn = 0
\end{equation}
Our tests indicate that $\lambda = 0.3$ is a reasonable choice which converges all studied molecules and, on average, speeds up the convergence by factor $4$ as compared to no mixing. As an example \figref{fig:norm_bf} shows the convergence using different mixing parameters for BF.

\section{Validation of qs$GW$}
\label{sect-qsGWvalidation}

In this section we will first confirm the internal consistency of the implemented qs$GW$ method. Second, we will benchmark, for a test set of 29 representative molecules, qs$GW$ versus $\Delta$CCSD(T) first IPs and experimental results of higher IPs of 5 organic molecules. Finally the results of qs$GW$ are compared to full self-consistency $GW$ (sc$GW$) literature results to validate the accuracy of qs$GW$ and to shed light on the internal features of the quasi-static approximation.

\subsection{qs$GW$ Internal Consistency}
\label{sect-internal}

We investigate the basis set dependence of the implemented qs$GW$ method and extract a basis set which yields a good tradeoff of computational cost and accuracy. Furthermore, we will show that the results of qs$GW$ are independent of the chosen initial functional in the initializing calculation.

\subsubsection{Basis Set Convergence}
\begin{figure}[bth]
  \includegraphics[width=0.99\columnwidth]{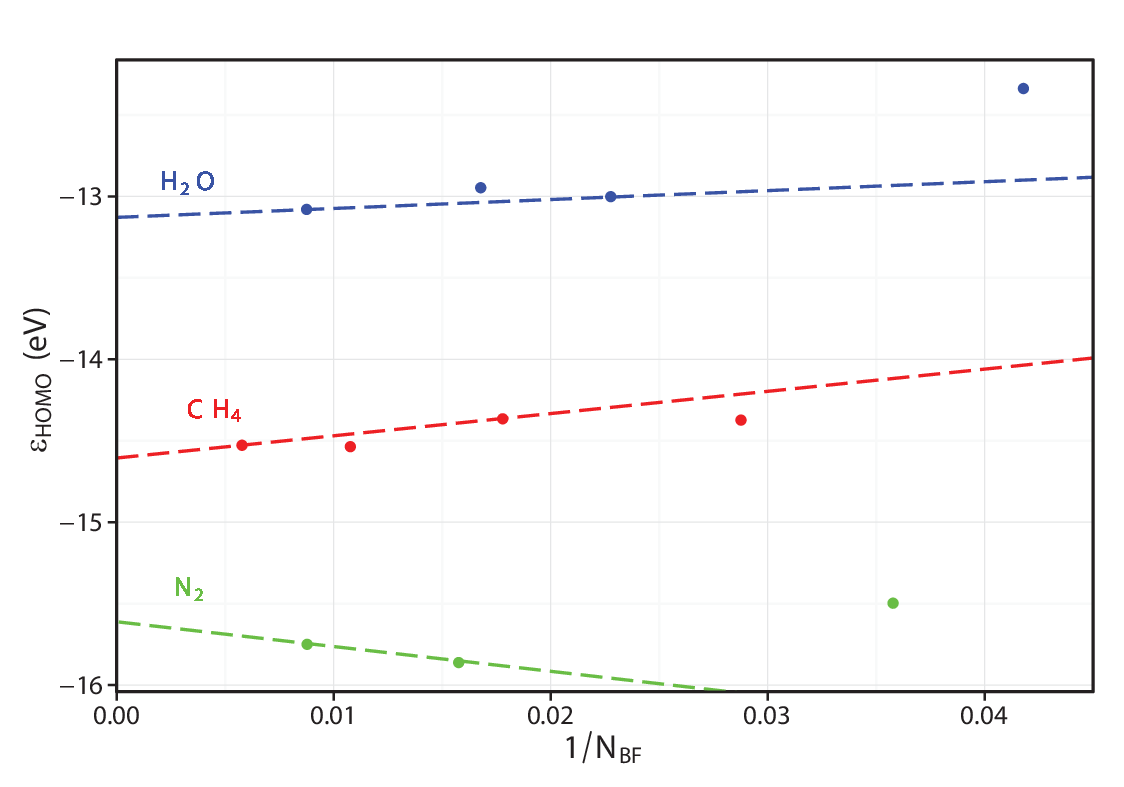}
   \caption{Convergence test of the qs$GW$-ionization potential with respect to the basis set (using, from left to right, the def2-QZVP, def2-TZVPP, def2-TZVP, def2-SVP basis sets) of water, nitrogen and methane. For nitrogen def2-TZVP and def2-TZVPP are identical. The intersection of the linear extrapolation of the def2-TZVP and def2-QZVP points with the ordinate gives an estimate for the complete basis set limit (CBS). }\label{fig:benchBasis}
\end{figure}

We start the study of the basis set dependence of the qs$GW$ IPs by a comparison for three small molecules, water, nitrogen and methane. The results obtained using the def2-SVP, def2-TZVP, def2-TZVPP and def2-QZVP basis set series are shown in \figref{fig:benchBasis}.\cite{turbomolebasis,tmri2} The results are plotted against the inverse of the size of the basis set. We take the ordinate offset of the linear extrapolation (the dashed lines in \figref{fig:benchBasis}) of the def2-TZVP and def2-QZVP as an estimate for the extrapolated complete basis set limit (CBS) result. This same approach has been used and tested in two of our previous studies for $G_0W_0$ comparing also to other basis sets.\cite{vansetten13,vansetten15gw100} It these studies it turned out have an estimated error of typically within 50 meV for $G_0W_0$.  Since the same physical quantities enter in qs$GW$ as in $G_0W_0$ it is reasonable to assume that the same holds for qs$GW$ as well. Indeed for these three molecules we observe the same rate of convergence as for $G_0W_0$.\cite{vansetten13} We therefore use this approach also for qs$GW$, at least as long as thoroughly tested extrapolation schemes as known for Hartree-Fock or coupled-cluster energies are not available. 

\begin{figure}[bth]
  \includegraphics[width=0.99\columnwidth]{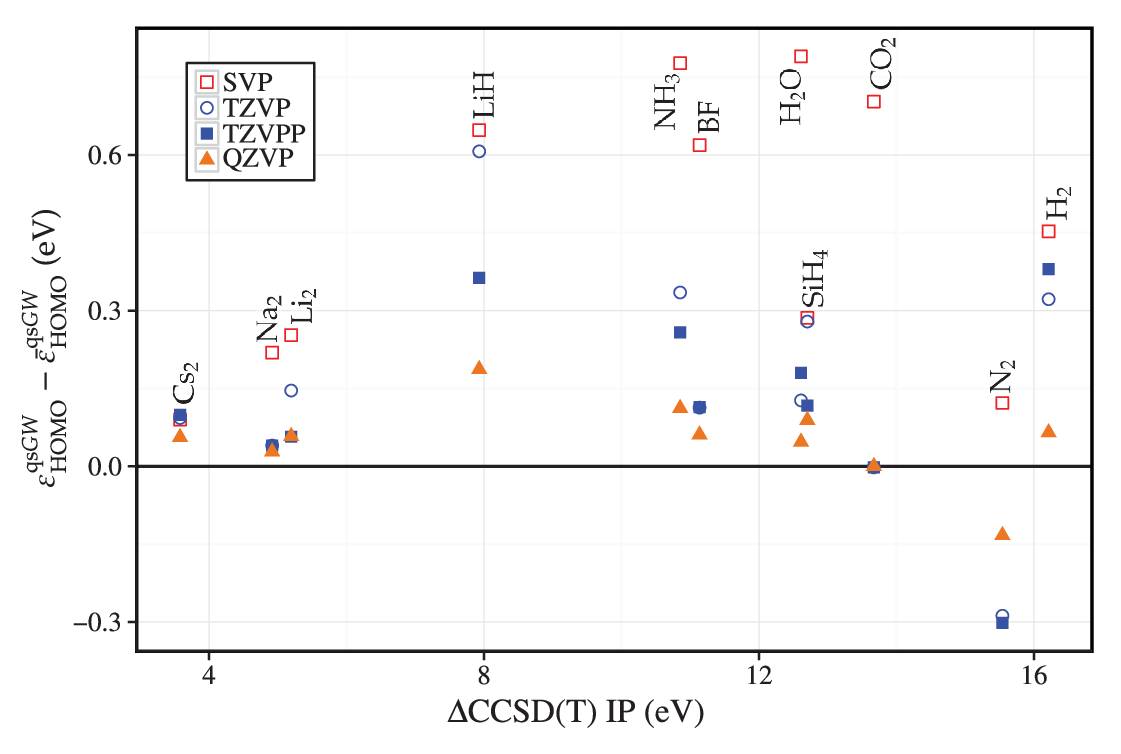}
   \caption{Change of the qs$GW$-ionization potential $\epsilon$ with increasing size of the basis set (using the def2-SVP, def2-TZVP, def2-TZVPP, def2-QZVP basis sets)\cite{turbomolebasis,tmri2} using the complet basis set limit (CBS) $\bar{\epsilon}$ as reference. The CBS is obtained from a linear interpolation over the inverse of the number of basis functions, see also \figref{fig:benchBasis}.}\label{fig:limitCBS}
\end{figure}

In a similar manner as shown in \figref{fig:benchBasis}, we have performed a survey over a larger subset of our test set of molecules and calculated the CBS limit for each individual molecule, the overall convergence behavior is indicated in \figref{fig:limitCBS}. For the def2-SVP basis set we find a maximum error larger then $0.8$~eV. Furthermore, the largest deviations are seen for systems with strong polar covalent bonds (H$_2$O, LiH, and NH$_3$). (This has already been observed for $G_0W_0$ in ref.~\citenum{vansetten13}). Our interpretation is that the def2-SVP basis set is not flexible enough to describe the high charge accumulation at one of the bond partners in ionic bonding situations. Therefore, the def2-SVP results have been excluded from the procedure to obtain the CBS limit.

Overall we find that the def2-TZVP basis shows a maximal deviation of roughly $0.4$~eV. Adding another set of polarization functions, the def2-TZVPP basis set, for most molecules the deviation drops below $0.3$~eV. Previously we have shown that for larger molecules the error actually is smaller.\cite{vansetten15gw100} Beyond def2-TZVP the quality of the $GW$ results mainly depends on the total number of basis functions, not necessarily the quality of the basis functions at the individual atoms. Moreover, the CCSD(T) results are available in the def2-TZVPP basis set, making those the ideal candidate for an accurate and unbiased comparison. We hence take the def2-TZVPP basis set as a reasonable compromise for the current study, with an uncertainty due to the basis set of maximally a few hundred meV. For really accurate calculations def2-QZVP basis set are however recommended. Using these for self-consistent calculations routinely however, require further parallelization of the response and GW routines in TURBOMOLE.

\subsubsection{Starting Point Dependence}

\begin{figure}[hbt]
   \includegraphics[width=0.99\columnwidth]{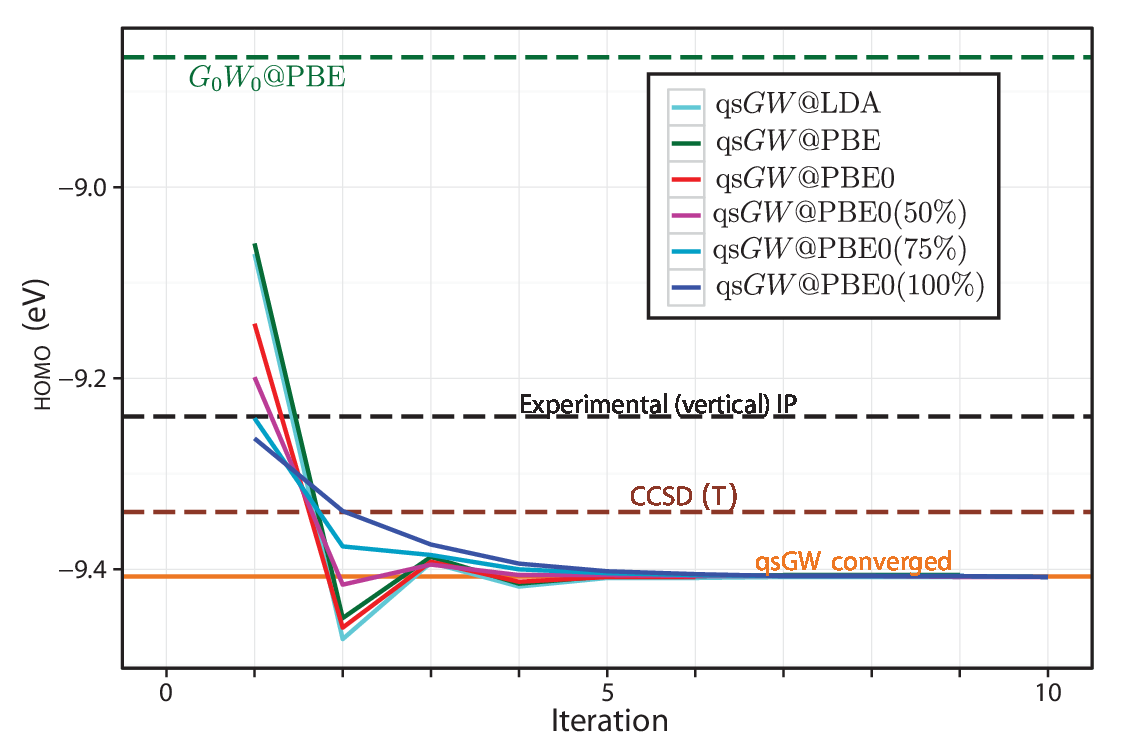}
  \caption{ The convergence of the HOMO energy for benzene with the qs$GW$ iterative cycle using the def2-TZVPP basis set. In this example the self-consistency cycle was initialized with a KS-DFT Green's function calculated employing LDA, PBE and PBE hybrid XC-functionals with an exact exchange contribution of 25\% (PBE0), 50\%, 75\%, and 100\%. For comparison also the $G_0W_0$, the experimental, and $\Delta$CCSD(T)\cite{krause15} results are shown.}
  \label{fig:evolution_Benzene}
\end{figure}

A major advantage of self-consistent $GW$ schemes is that the fixed-point found in the iteration scheme is for the commonly used starting points (largely) independent of the initialization.\cite{tandetzky12arx} This is true also for qs$GW$, as is illustrated in \figref{fig:evolution_Benzene}. The convergence of the HOMO energy with increasing cycle number starting from different Kohn-Sham XC-functionals is shown for the example of benzene. In this work we explore the LDA,\cite{lda} PBE,\cite{tmf8} PBE0,\cite{tmf11} and PBE0(X\%) (PBE0 with X \% exact exchange).  The iterative procedure converges into exactly the same solution, independent of the starting point.

\subsection{First Ionization Potential: comparison to $\Delta$CCSD(T)} 
\label{sect-1stIPqs}

As a first quantitative test we calculate the ionization energies/potentials (IP) for a test set of 29 molecules ranging from H$_2$ to tetrathiafulvalene.\cite{gw29} We focus on IPs since (i) they are an important indicator to understand charge transfer processes (ii) experimental reference data is available and (iii) one has access to results using more accurate theories (at least for small size molecules). The first IP is trivially extracted from the calculated data being the (negative) energy of the highest occupied molecular orbital (HOMO). These, and all following results have been calculated using the def2-TZVPP basis set.\cite{turbomolebasis,tmri2}

The $G_0W_0$ results have been calculated initialized from DFT using the PBE functional, denoted $G_0W_0$@PBE. In the calculation of the correlation part of the self-energy the positive infinitesimal $\eta$ was chosen so that all (orbital-)energies are converged within 1~meV. Typically $\eta=1$~meV was used. For all calculations the RI approximation was used. It has been shown previously that the errors introduced by RI, applying the standard auxiliary basis functions,\cite{tmri2} are, except for very small systems like He, Ne, and H$_2$, below than 100~meV.\cite{vansetten15gw100} The parameters for the qs$GW$ calculations were chosen to converge the pole-positions within 1~meV.

The obtained IPs are compared to results obtained by employing the coupled-cluster method in the CCSD(T) approximation,\cite{krause15} using the same atomic structures and same basis set as were used in the $GW$ calculations (def2-TZVPP).

\begin{figure}[!htb]
   \includegraphics[width=0.95\columnwidth]{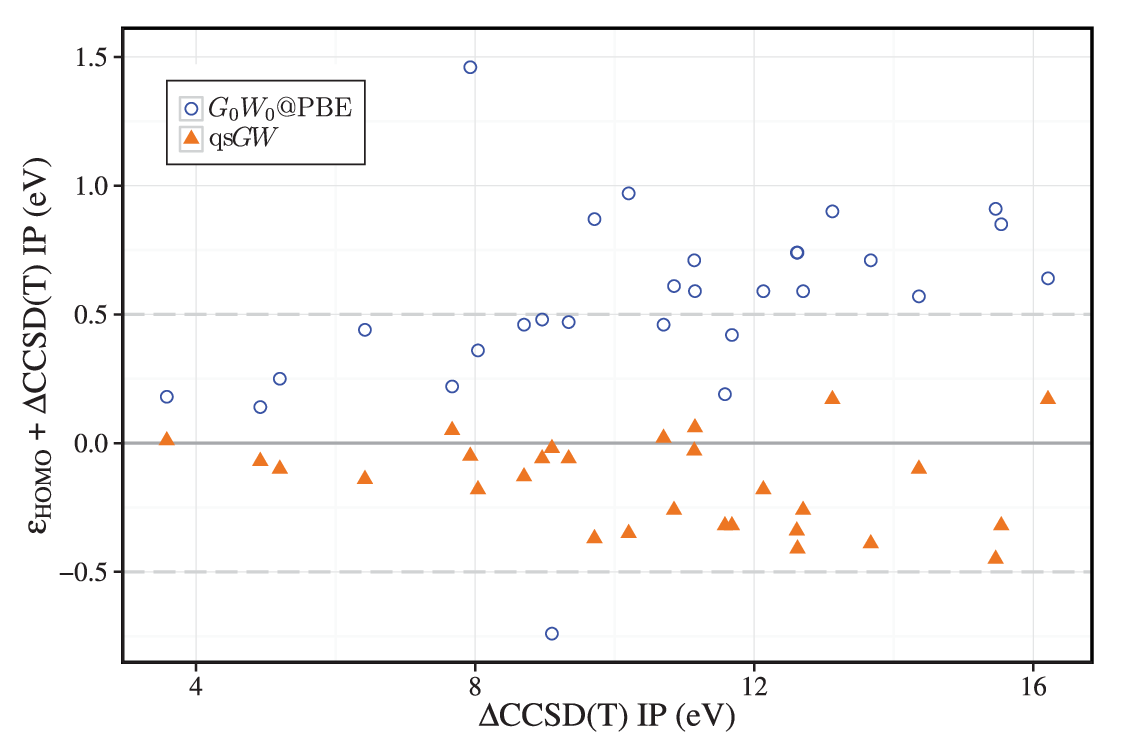} 
  \caption{ Deviations of the HOMO energies using $G_0W_0$@PBE and qs$GW$ with IPs from $\Delta$CCSD(T)\cite{krause15} calculations. The corresponding numerial data is provided in \tabref{tab:homos_complete} and statistical evaluation of the deviations is shown in \tabref{tap:statisticsCC_IP_qsgw}.}\label{fig:distanceHomo}
\end{figure}

The deviations in the calculated HOMO energies using the different flavors of $GW$ to $\Delta$CCSD(T) IPs are displayed in \figref{fig:distanceHomo}. Together with the experimental (vertical) IPs the calculated IPs are reported in \tabref{tab:homos_complete}. The data shows an improved agreement of qs$GW$ with $\Delta$CCSD(T) in comparison to the $G_0W_0$@PBE results, by up to $1.41eV$. The mean absolute error (MAE) improves by $0.40$~eV, see \tabref{tap:statisticsCC_IP_qsgw}, from the single-shot $G_0W_0$ to the self-consistent qs$GW$ estimates.  Furthermore, the increasing trend in the discrepancy is cleared within qs$GW$.

\begin{table}[!htbp]
\begin{center}
\begin{tabular}{lrrrr}
\hline\hline
molecule & exp. & $\Delta$CCSD(T)\cite{krause15} & $G_0W_0$ & qs$GW$ \\ \hline
H$_2$ & 15.42 & 16.21 & 15.57 & 16.04 \\ 
Li$_2$ & 5.11 & 5.20 & 4.95 & 5.30 \\ 
Na$_2$ & 4.89 & 4.92 & 4.78 & 4.99 \\ 
Cs$_2$ & 3.70 & 3.58 & 3.40 & 3.57 \\ 
F$_2$ & 15.70 & 15.46 & 14.55 & 15.91 \\ 
N$_2$ & 15.58 & 15.54 & 14.69 & 15.86 \\ 
BF & 11.00 & 11.14 & 10.43 & 11.17 \\ 
LiH & 7.90 & 7.93 & 6.47 & 7.98 \\ 
CO$_2$ & 13.78 & 13.67 & 12.96 & 14.06 \\ 
H$_2$O & 12.62 & 12.61 & 11.87 & 12.95 \\ 
NH$_3$ & 10.85 & 10.85 & 10.24 & 11.11 \\ 
SiH$_4$ & 12.82 & 12.70 & 12.11 & 12.96 \\ 
SF$_4$ & 12.30 & 12.62 & 11.88 & 13.03 \\ 
Au$_2$ & 9.50 & 9.10 & 9.84 & 9.12 \\ 
Au$_4$ & 8.60 & 7.67 & 7.45 & 7.62 \\ 
methane & 14.35 & 14.36 & 13.79 & 14.46 \\ 
ethane & 12.00 & 13.12 & 12.22 & 12.95 \\ 
propane & 11.51 & 12.13 & 11.54 & 12.31 \\ 
butane & 11.09 & 11.58 & 11.39 & 11.90 \\ 
isobutane & 11.13 & 11.68 & 11.26 & 12.00 \\ 
ethylene & 10.68 & 10.70 & 10.24 & 10.68 \\ 
acetone & 9.70 & 9.71 & 8.84 & 10.08 \\ 
acrolein & 10.11 & 10.20 & 9.23 & 10.55 \\ 
benzene & 9.24 & 9.34 & 8.87 & 9.40 \\ 
naphthalene & 8.09 & 8.04 & 7.68 & 8.22 \\ 
thiophene & 8.85 & 8.96$^a$ & 8.48 & 9.02 \\ 
benzothiazole & 8.75 & 8.70$^a$ & 8.24 & 8.83 \\ 
1,2,5-thiadiazole & 10.11 & 10.09$^a$ & 9.65 & 10.18 \\ 
tetrathiofulvalene & 6.72 & 6.42$^a$ & 5.98 & 6.56 \\ 
\hline\hline
\end{tabular}
\mbox{}\\
$^a$ This work.
\end{center}
\caption{The calculated (minus) HOMO energies from qs$GW$ and $G_0W_0$ (initialized from DFT employing the PBE functional using the def2-TZVPP basis set) as well as experimental (vertical) Ionization Potentials and estimates from $\Delta$CCSD(T).\cite{krause15} All values are in eV and all calculated results are obtained wihtin the def2-TZVPP basis set.}
\label{tab:homos_complete}
\end{table}

\begin{table}[!htb]
\centering
\begin{tabular}{lrr}
\hline\hline
 & $G_0W_0$ & qs$GW$ \\ \hline
ME & 0.57 & -0.16 \\ 
MAE & 0.59 & 0.19 \\ 
$\sigma^2$ & 0.08 & 0.02 \\ 
MaxAE & 1.49 & 0.45 \\ 
MinAE & 0.14 & 0.01 \\ 
\hline\hline
\end{tabular} \caption[IP statistics of qs$GW$]{Statistical measures of the difference of the calculated IP from $G_0W_0$ and qs$GW$ to the reference $\Delta$CCSD(T) IP cumulated over the test set. All values are in eV.}\label{tap:statisticsCC_IP_qsgw}
\end{table}

\subsection{Higher Ionization Potentials: comparison to sc$GW$ and experiment}
\label{sect-higherIPqs}

\label{sect-qsvssc}

In this section we discuss the accuracy of qs$GW$ for the calculation of higher IPs. The higher IPs are difficult to access via CCSD(T). For the first ionization energy the total energy difference between the neutral and cationic ground state has to be calculated. Calculating the latter correctly is already delicate. For the higher ionization energies excited states of the cation have to be calculated. This makes these calculations very cumbersome (sometimes impossible). We therefore fall back to experimental values as references.\cite{naphthalene}

We extended our test with five organic molecules that are candidates for optical devices and for which also experimental data is available. These molecules have already been investigated with sc$GW$ by Caruso et al.\cite{caruso12prb, caruso13prb} We will compare the $G_0W_0$, qs$GW$, and spectra  to experimental results. The results are shown for naphthalene in this section, the results on three more molecules (Thiophene, Benzothiazole, and 1,2,3-Thiadiazole) are available in the supplementary information.

In \figref{fig:scGW_naphthalene} higher IPs using $G_0W_0$, qs$GW$ as well as sc$GW$ are compared to the experimental (vertical) IPs for naphthalene. The $G_0W_0$ and qs$GW$ methods show similar behavior as described in the previous section. Furthermore, IPs from qs$GW$ and sc$GW$ give similar agreement with experiment. qs$GW$ clearly outperforms sc$GW$ for the lowest ionization energies but for the higher ionization energies sc$GW$ becomes better. Both self-consistent methods, qs$GW$ and sc$GW$, remove most of the energy dependence in the error of $G_0W_0$. Actually, we find a rigid shift of about $0.5$~eV for all energy levels between qs$GW$ and sc$GW$ for nearly all molecules investigated. A similar picture is observed for the three further molecules where we compare the higher ionization energies between qs and sc$GW$ (see Supplementary information). For these rather similar aromatic systems we observe in all cases a shift $\sim0.5$~eV. Whether this holds for a broader class of molecules needs further systematic investigation, especially since the molecules in this comparison exhibit very similar bonding.

\begin{figure}[bth]
  \includegraphics[width=0.95\columnwidth]{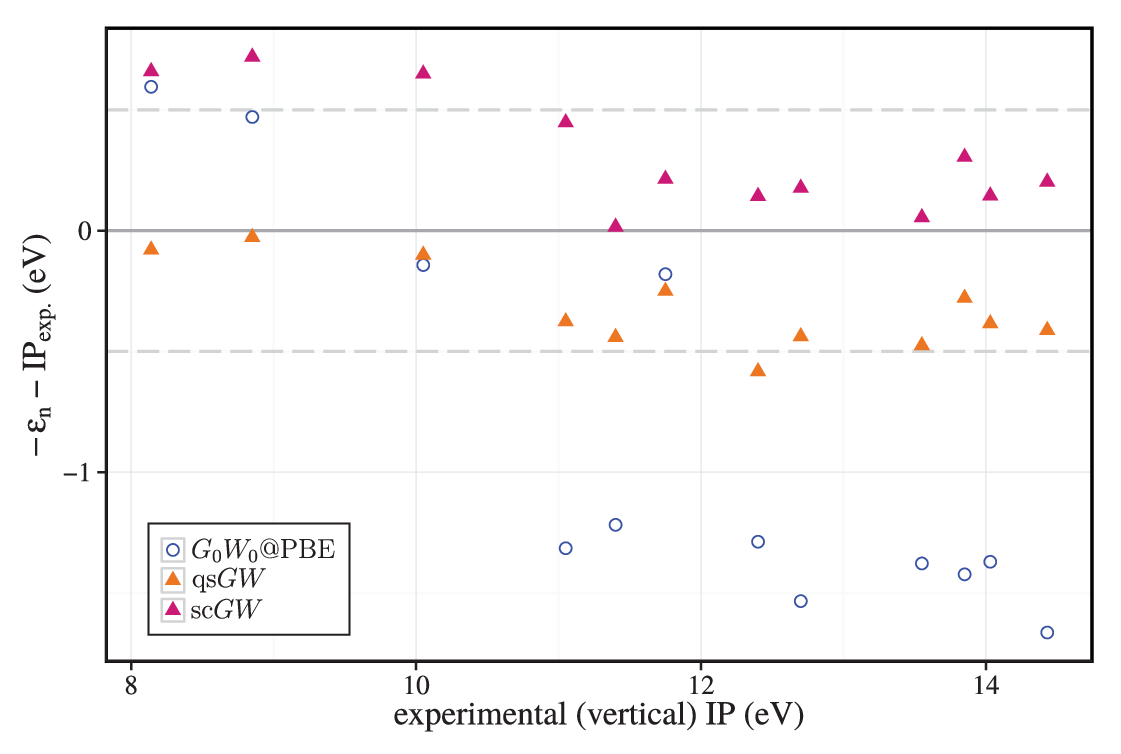}
   \caption{Deviation of the QP-energies betweeb $\GoWo$, qs$GW$ and sc$GW$\cite{caruso12prb} (FHI-AIMS) from experimental ionization energies for naphthalene. The dashed line indicates the deviation of $\pm0.5$~eV.} \label{fig:scGW_naphthalene}
\end{figure}

The shift can be understood by recalling the split of the Green's function into a quasi-particle part and an incoherent part 
\begin{equation}
G = ZG^{QP} + \bar{G} 
\end{equation}
In qs$GW$ the $Z$-factor is unity and the incoherent part $\bar{G}$ is neglected, while in the sc$GW$ both are taken into account. We suggest that $\bar{G}$ is only weakly energy dependent, so that its contribution $\bar{\Sigma}= i\bar{G}W$ to the self-energy is also a relatively flat function of energy. The current data, and the results by Koval {\em et al.}\cite{koval14}, show that for molecules the effect of the quasi particle approximation however has the opposite sign as in solids. Vanschilfgaarde and coworkers argue that calculating the response from the full green function, as in scGW, would lead to under screening and hence overestimation of the ionization energies.\cite{vanschilfgaarde06} This is indeed what one would  conclude from results on the homogeneous electron gas.\cite{holm98} In the present results on molecules scGW underestimates and qsGW overestimates. A systematic study aiming to understand this inversion including many more molecules, exhibiting different kinds of bonding is currently in progress.

\subsection{Comparison to qs$GW$ Literature Results}

Recently Koval {\em et al.} also compared qs$GW$ and sc$GW$ for a set of molecules.\cite{koval14} Their work employs a spectral function technique on an equidistant energy grid to calculate the convolution of $G$ and $W$. Their method 'A' is the same as the approach used in this work, Bruneval's implementation in molgw,\cite{bruneval/atom} and the original implementation by Faleev {\em et al.}\cite{vanschilfgaarde04prl} for solids.\cite{notekoval}

To make a definite test for the agreement between the three different implementations of qs$GW$ for molecules, we calculate the helium atom using the basis set employed in the works by Koval {\em et al.} and Bruneval, the correlation-consistent basis sets.\cite{dunning:1989A} The results are compared in \tabref{litcomp}. We observe an excellent agreement with Brunevals results at the meV level. In contrast, we observe a deviation to the 'A' method of Koval {\em et al.} that is quite substantial, in the tens of meV range. For comparison \tabref{litcomp} also lists the results from the approximate 'B' method. 

\begin{table}[!htbp]
\begin{center}
\begin{tabular}{lcccc}
\hline\hline
basis&  qs$GW$ B\cite{koval14} & qs$GW$ A\cite{koval14} & qs$GW$\cite{bruneval/atom}& qs$GW$ (this work)\\ 
\hline
cc-pVDZ & 24.346 & 24.350	 & 24.359 & 24.359\\
cc-pVTZ & 24.554 & 24.340 & 24.320 & 24.320\\
cc-pVQZ & 24.668 & 24.751	 & 24.766 & 24.767\\
cc-pV5Z & 24.705 & 24.799	 & 24.825 & 24.826\\
\hline\hline
\end{tabular}
\end{center}
\caption{HOMO energies for atomic helium from qs$GW$, literature comparison. All values are in eV.}
\label{litcomp}
\end{table}

\subsection{Densities and Dipole Moments.}
\label{sect-density}

A self-consistent solution of the $GW$ equations introduces corrections in the spatial shape of the QP-orbitals in addition to shifts of the  pole positions of the Green's function. Hence, corrections to the ground state density as compared to the reference density can arise.

\begin{figure}[!htb]
  \includegraphics[width=0.95\columnwidth]{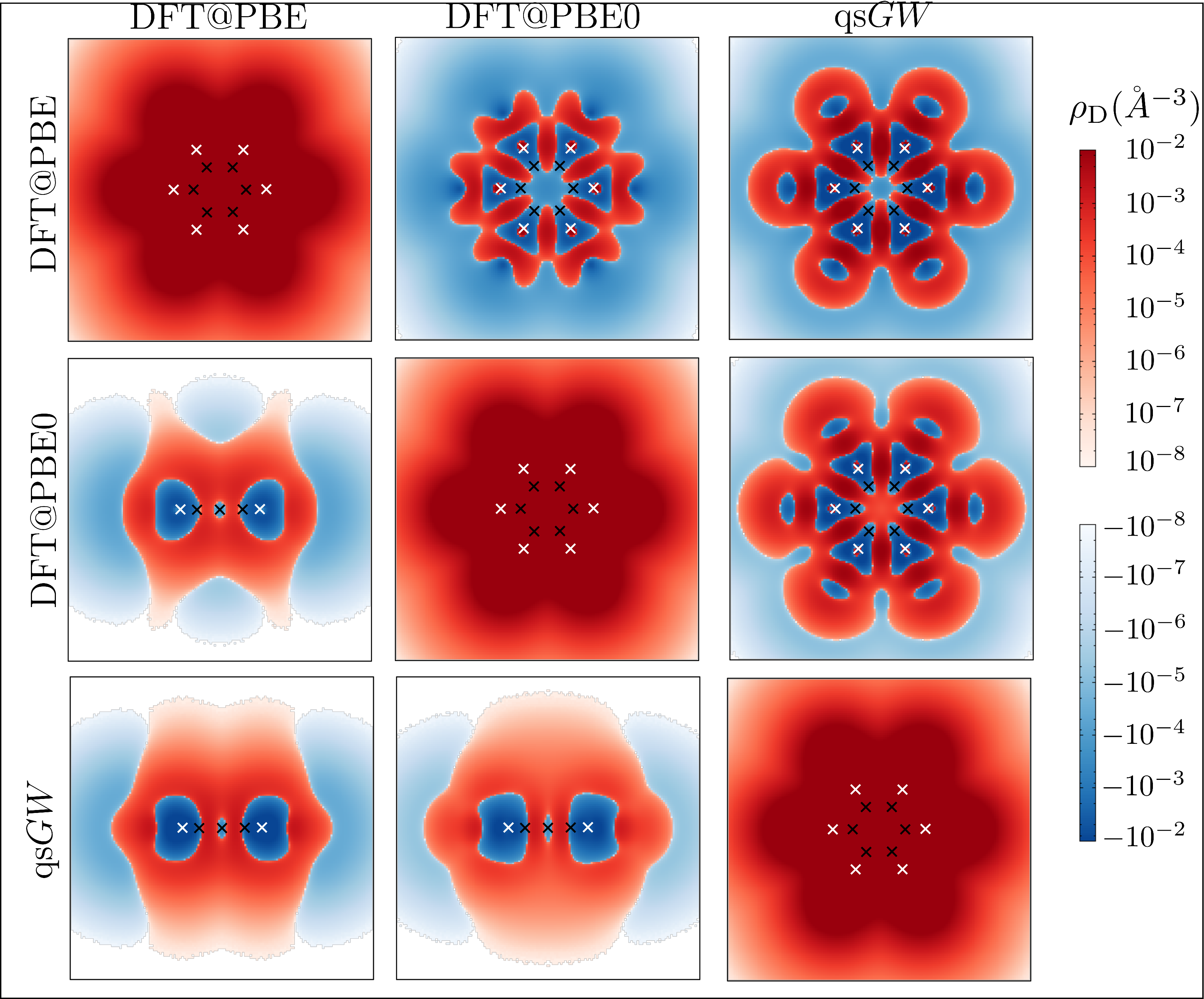} \caption{Diagonal: Ground state electron density of benzene as obtained from DFT(PBE), DFT(PBE0) and qs$GW$. Off-diagonal upper right: Difference in the ground state electron densities: PBE-PBE0, PBE-qs$GW$, and PBE0-qs$GW$, horizontal cut. Off-diagonal lower left: Same differences, vertical cut. The positions of the atoms are indicated by the black (carbon) and white (hydrogen) crosses. The results from the qs$GW$ calculation show a slightly stronger localized density (blue area) around the H-C paires in comparison to the DFT(PBE)  calculations. }
  \label{fig:dens_benzene}
\end{figure}

Figure~\ref{fig:dens_benzene} shows the differences between the calculated electron densities of benzene for DFT(PBE), DFT(PBE0) and qs$GW$, showing that the qs$GW$ orbitals are slightly more localized. This localization is cause by the cancelation of a large part of the the spurious self-interaction of the approximate DFT functionals. The part of the self interaction that is caused by self-screening is not canceled in qs$GW$.\cite{nelson07} Removing this as well would require the inclusion of more diagrams.\cite{aryasetiawan12}  

\begin{figure}[!htb]
  \includegraphics[width=0.95\columnwidth]{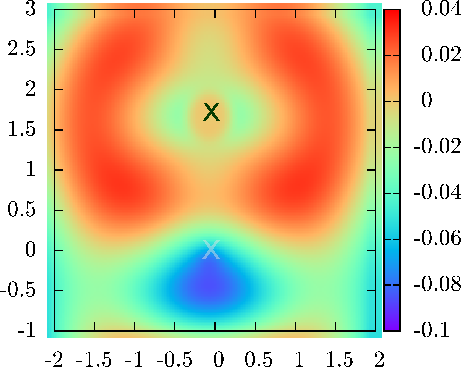}
   \caption{Difference between the calculated DFT (PBE) density and the qs$GW$ density normalized on the (initial) DFT density for hydrogenflouride. The fluor atom is denoted by a green cross, the hydrogen atom by a grey cross. }
   \label{fig:dens_dft}
\end{figure}

The enhanced tendency towards localization displays clearer looking at an ionic molecule like HF, see \figref{fig:dens_dft}; we observe a more localized density around the fluoride atom. Furthermore, the amount of charge on the non-bonding side of the hydrogen atom is slightly reduced. The trend seen here for qs$GW$ agrees with the observations of Caruso {\em et al.} for sc$GW$.\cite{caruso13prb}

An experimentally easily accessible observable characterizing the ground state density is the dipole moment. The comparison is given in \tabref{tab:dipoles}. To also be able to compare to sc$GW$ we will present here the systems used by Caruso {\em et al.}. 

While HF typically overestimates the dipole moment of the dimers, DFT typically slightly underestimates it. In the range of hybrid functionals the best dipole moments are obtained from the PBE0 functional. sc$GW$ performs comparable to PBE0. In contrast, qs$GW$ yields better agreement with an overall mean absolute error (MAE) of $0.03$~Debye. To make a definitive statistically sound statement discriminating between the qs$GW$ and sc$GW$ is however not possible due to the small number of systems.  In conclusion, our results support the general impression that self-consistent $GW$, is a promising tool to investigate the charge-transfer in molecular interfaces and other (nano-scale) hetero-structures.\cite{caruso14}

\begin{table}[!htb]
\begin{tabular}{lcccccc}
\hline\hline
 & LiH  & HF  & LiF  & CO  & ME & MAE  \\ 
\hline
Exp. & 5.88 & 1.82 & 6.28 & 0.11 & & \\
PBE & 5.60 & 1.80 & 5.99 & 0.24 & -0.12 & 0.18\\
PBE0(25\%) & 5.77 & 1.85 & 6.18 & 0.11 & -0.05 & 0.06\\
PBE0(75\%) & 6.01 & 1.93 & 6.45 & 0.14 & 0.11 & 0.11\\
PBE0(100\%) & 6.10 & 1.97 & 6.54 & 0.26 & 0.20 & 0.20\\
HF & 6.03 & 1.95 & 6.50 & 0.26 & 0.16 & 0.16\\
sc$GW$\cite{caruso12prb} & 5.9 & 1.85 & 6.48 & 0.07 & 0.05 & 0.07\\
qs$GW$ & 5.83 & 1.84 & 6.29 & 0.07 & -0.02 & 0.03\\
\hline\hline
\end{tabular} \caption{Comparison between experimental\cite{dipolemoments} and theoretical dipole moments (Debye), from PBE, PBE0, PBE0(75\%), PBE0(100\%), HF, sc$GW$\cite{caruso12prb} (FHI-AIMS) and qs$GW$. For all calculations the experimental equilibrium bond length was considered, the calculations of this work are performed in the def2-TZVPP basis-set.}\label{tab:dipoles}
\end{table}

Comparing the differences between sc and qs $GW$ for on one hand the ionization energies and on the other hand the dipole moments we observe a different degree of agreement. The dipole moments tend to agree better between the two methods than the ionization energies. We assume this to be related to the rigid shift we observed between two respective spectra. 
					
\section{Comparison of qs$GW$ and $G_0W_0$}
\label{sect-g0w0comparison}

Given the qs$GW$ results for the IPs from the previous sections, we next evaluate next how well the traditional $G_0W_0$ ($G_0W_0$0th) is capable of reproducing them if one employs an improved DFT starting point. One can do so by using parametrized functionals derived from the PBE0 hybrid functional. In addition, the $G_0W_0$ 2nd order approach\cite{kaplan15gwso} ($G_0W_0$2nd), which takes into account off-diagonal elements of $\Sigma$ in the QP-equation and hence an influence of orbital corrections, is also tested.

\subsection{First Ionization Potentials}

\begin{figure}[!htb]
   \includegraphics[width=0.95\columnwidth]{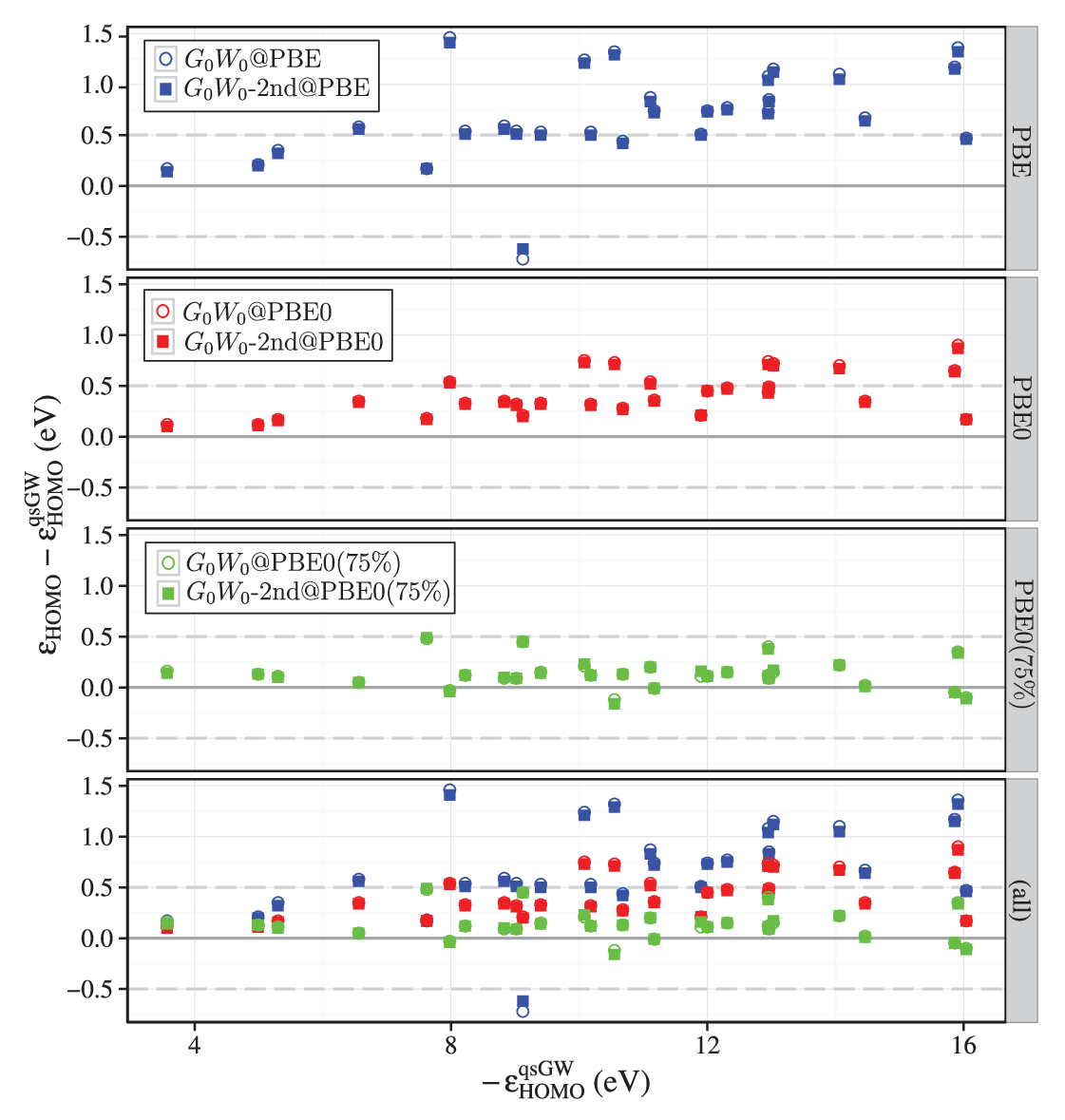} 
  \caption{ Deviation of HOMO energies obtained from  $G_0W_0$0th and $G_0W_0$2nd to qs$GW$. Results are shown for calculations  initialized from DFT, employing PBE and PBE hybrid XC-functionals with an exact exchange contribution of 25\% (PBE0) and 75\% (PBE0(75\%)). On average best agreement is achieved employing the PBE0(75\%) starting point, see \tabref{tab:statistics_scnd_qsgw}.}
  \label{fig:distance_G0W0_qsGW}
\end{figure}

The HOMO energies obtained from $G_0W_0$0th@PBE (traditional $G_0W_0$ with a PBE starting point) have a mean absolute error 0.75~eV and show a clear correlation between the error and the actual value. With increasing energy, the error, as compared to qs$GW$, systematically increases. Employing a PBE0 starting point improves the overall agreement with qs$GW$ down to a mean absolute deviation of $0.42$~eV, see \figref{fig:distance_G0W0_qsGW} and \tabref{tab:statistics_scnd_qsgw}. Best agreement is achieved employing the PBE-hybrid based starting point employing  75\% exact exchange. For this starting point, the correlation between the error and ionizationenergy disappears. Our data also confirms a trend that already appeared in our earlier work: $G_0W_0$0th and $G_0W_0$2th give similar results. This suggests that off-diagonal elements of $\Sigma$ give negligible contributions to the QP-energies.\cite{kaplan15gwso}

\begin{table}[!htb]
\centering
\begin{tabular}{l c c c c  c c}
\hline\hline
\multirow{2}{*}{$G_0W_0$} & \multicolumn{2}{c}{@PBE} & \multicolumn{2}{c}{@PBE0} & \multicolumn{2}{c}{@PBE0(75\%)} \\
& ~0th~ & 2nd & ~0th~ & 2nd & ~0th~ & 2nd\\ 
\hline
ME & 0.70 & 0.67 & 0.42 & 0.41 & 0.11 & 0.19 \\ 
MAE & 0.75 & 0.72 & 0.42 & 0.41 & 0.18 & 0.26 \\ 
$\sigma^2$ & 0.20 & 0.18 & 0.05 & 0.04 & 0.05 & 0.23 \\ 
MaxAE & 1.46 & 1.41 & 0.90 & 0.87 & 0.85 & 2.41 \\ 
MinAE & 0.17 & 0.14 & 0.12 & 0.10 & 0.01 & 0.01 \\ 
\hline\hline
\end{tabular} \caption[$G_0W_$ first IPs vs. qs$GW$]{Statistical measures over the data from \figref{fig:distance_G0W0_qsGW}.  Evaluated is the difference of the calculated $G_0W_0$ and $G_0W_0$-2nd HOMO energies to the qs$GW$ HOMO cumulated over the test set. Three different DFT based starting points have been  employed. On average, best agreement is achieved with the  PBE hybrid XC-functionals with a contribution of exact exchange of 75\% as the starting point (PBE0(75\%)). All values are in eV.}\label{tab:statistics_scnd_qsgw}
\end{table}

\subsection{Higher Ionization Potentials}

\begin{figure}[bth]
\includegraphics[width=0.95\columnwidth]{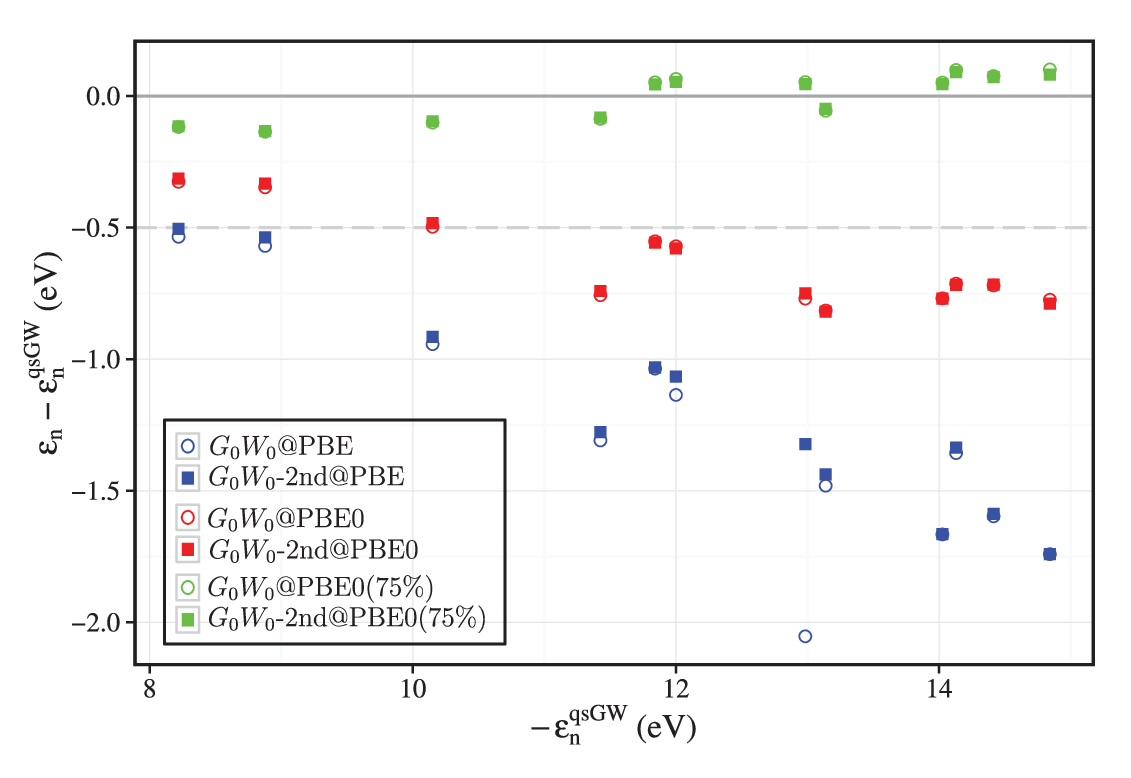}
   \caption{Comparison of the QP-energies from $G_0W_0$0th and $G_0W_0$2nd with qs$GW$ for naphthalene. Three different DFT based starting points employing the PBE, and the PBE hybrid functional with 25\% (PBE0) and with 75\% exact exchange (PBE0(75\%)), were chosen. The best agreement, with qs$GW$, yields the calculation with the PBE0(75\%) starting point. }
   \label{fig:scnd_naphthalene}
\end{figure}

Figure~\ref{fig:scnd_naphthalene} displays the QP-energies from $G_0W_0$0th and $G_0W_0$2th relative to qs$GW$ for different PBE0 based starting points for naphthalene. In our example we find best agreement, with qs$GW,$ employing the starting point with a higher contribution of exchange (PBE0(75\%)). Furthermore, we see in the calculation initialized from PBE0(75\%) no energy dependent trend in the error, while the trend persists in both $G_0W_0$0th and $G_0W_0$2nd employing a PBE or PBE0 starting point.

\section{Validating Partial Self-Consistency}
\label{sect-partialsc}

Even though computationally more affordable than typical coupled cluster approaches, qs$GW$ is still too expensive for systematic screenings over large ensembles of intermediate sized molecules. Thus motivated, in this section we test simplified flavors of qs$GW$. Our aim is to find an approximation scheme that captures the dominant contributions included in qs$GW$ and hence is capable of reproducing results of qs$GW$, at a lower computational cost. 

\subsection{qs$GW$ with Fixed Orbitals: ev$GW$}
Our studies of the charge density for qs$GW$ suggest that orbital updates tend to be small. Therefore, we next ignore orbital updates in the self-consistency cycle and take into account only the update in the 
pole positions: eigenvalue only $GW$ (ev$GW$).\cite{olevano11prb}

\subsubsection{Method ev$GW$}

Within the methodology described in section~\ref{sec:qsgw} turning qs$GW$ into ev$GW$ is straight forward, one only needs to suppress the update in the orbitals between consecutive qs$GW$ iterations. In other words the machinery operates solely with the initial set of QP-orbitals $\psi^{(0)}(\br)$ and the orbital update from \ref{eq:ks2qp} is skipped. In all equations the orbitals $\psi^{(i)}(\br)$ are replaced with $\psi^{(0)}(\br)$. Furthermore. our treatment of the quasiparticle equation in ev$GW$ neglects off-diagonal elements of $\Sigma$. Finally we discard the static approximation and restore in the QP-equation the energy-dependent self-energy $\Sigma$. Then, for each pole the QP-equation is solved self-consistently. Computationally this introduces the benefit that the Coulomb exchange integrals, see \ref{eq:exchange}, do not need to be re-evaluated. For instance, the full exchange part of the self-energy $\Sigma_{\text{x}}$, see \ref{eq:exchange}, needs only one single evaluation. 

\subsubsection{Comparison to qs$GW$ - First IPs}

\begin{figure}[!htb]
   \includegraphics[width=0.95\columnwidth]{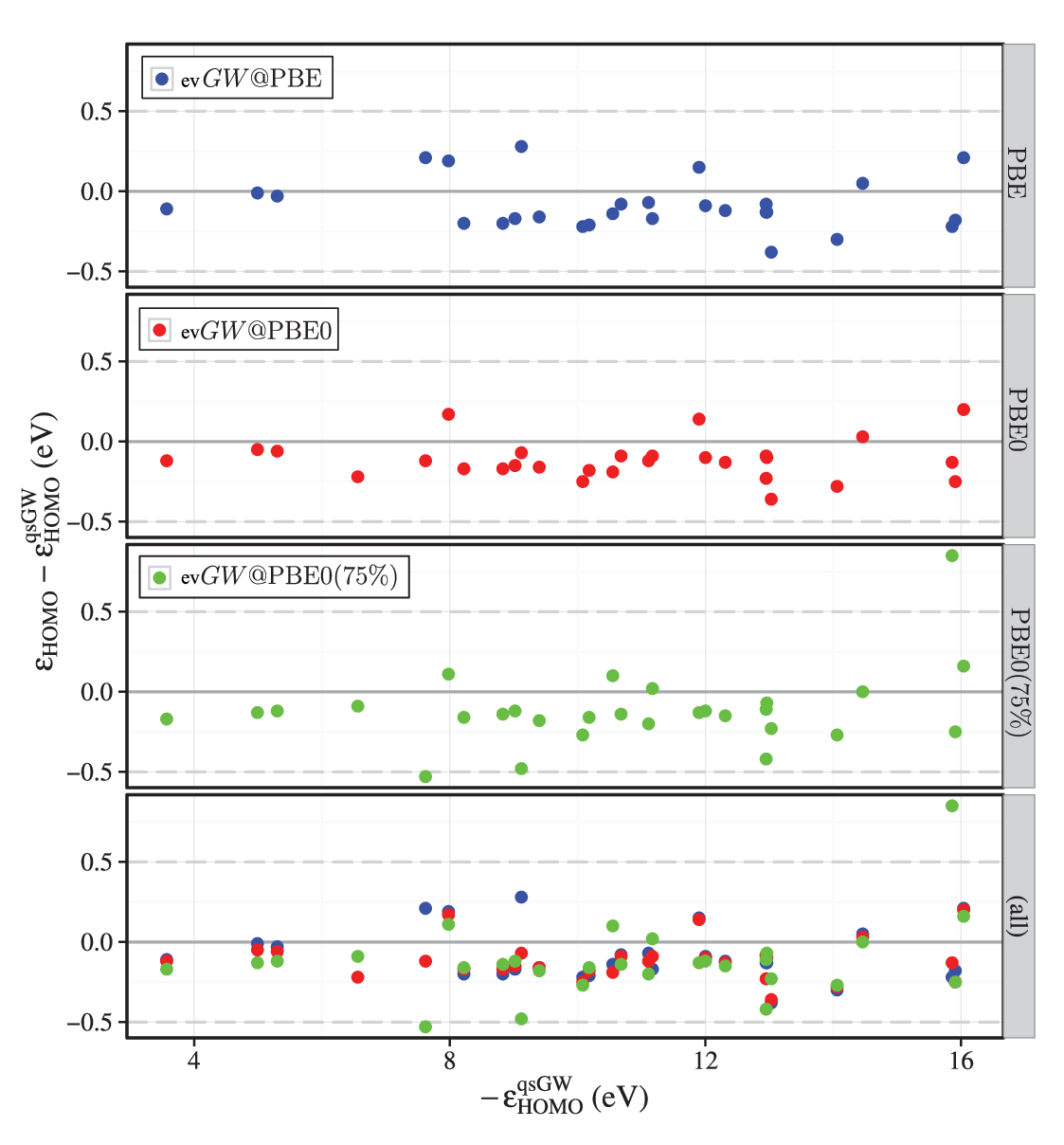} 
  \caption{Comparison of HOMO energies using ev$GW$ initialized from DFT calculations with the qs$GW$ HOMO energies for the test set. The ev$GW$ starting point was constructed from PBE and PBE-hybrid calculations. The results of ev$GW$ show only weak dependence on the amount of  exact exchange in the functional of the initial DFT calculation.}\label{fig:distance_evgw_qsGW} 
\end{figure}

For the set of molecules specified in \tabref{tab:homos_complete}, the differences between the ev$GW$ estimates of first IPs and the qs$GW$ results are shown in \figref{fig:distance_evgw_qsGW} (see \tabref{tab:statistics_evgw_start} for the statistical evaluation). Because ev$GW$ is not self-consistent in the orbitals there is a residual starting point dependence, which is clearly visible in \figref{fig:distance_evgw_qsGW}. It is however significantly reduced as compared to, e.g., $G_0W_0$. In contrast to the findings from the $G_0W_0$ before we find best overall agreement with the qs$GW$ by employing the PBE0 functional (and not anymore the PBE0(75\%) hybrid with 75\% exact exchange). Even for pure PBE the overall agreement with qsGW is better than with PBE0(75\%). 

\begin{table}[!htb]
\centering
\begin{tabular}{lrrrrr}
\hline\hline
  ev$GW$  & @PBE &  @PBE0 &   @PBE0(75\%)   \\
\hline
ME & 0.09 & 0.12 & 0.12 \\ 
MAE & 0.17 & 0.15 & 0.20 \\ 
$\sigma^2$ & 0.03 & 0.02 & 0.06 \\ 
MaxAE & 0.39 & 0.36 & 0.85 \\ 
MinAE & 0.01 & 0.03 & 0.00 \\ 
\hline\hline
\end{tabular} \caption[ev$GW$ first IPs vs. qsGW]{Statistical evaluation of the data from \figref{fig:distance_evgw_qsGW}. The deviation from calculated ev$GW$ HOMO energies (from different PBE and PBE-hybrid starting points) to the qs$GW$ HOMO cumulated over the test set. With ev$GW$ we find best agreement with qs$GW$ if the PBE0 starting point is employed. All values are in eV.}\label{tab:statistics_evgw_start}
\end{table}

\subsubsection{Comparison to qs$GW$ - Higher IPs}
\begin{figure}[bth]
  \includegraphics[width=0.95\columnwidth]{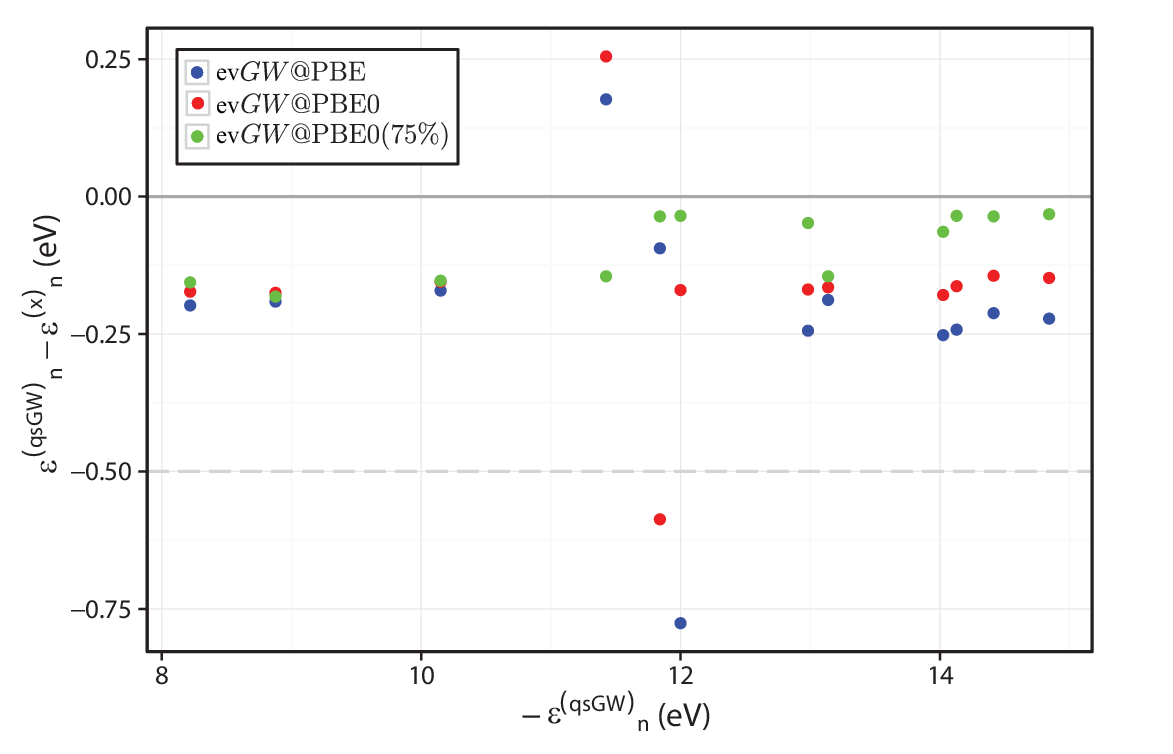}
   \caption{Deviation of the QP-energies from ev$GW$ from different PBE and PBE-Hyrbid starting points to  qs$GW$ for naphthalene.}\label{fig:evGW_naphthalene}
\end{figure}

A comparison of QP-energies from ev$GW$ and qs$GW$ is given in \figref{fig:evGW_naphthalene} for naphthalene. The results confirm the findings for the first IPs. The starting point dependence is strongly reduced and we find that the trend in the energy dependence of the error, which was present in $G_{0}W_0$, is not present anymore for all starting points. All in all, self-consistency in the poles by it self already produces very good agreement with qs$GW$ QP-energies for all starting points.

\subsubsection{Comparison to $\Delta$CCSD(T) - first IPs}
\begin{figure}[!htb]
   \includegraphics[width=0.95\columnwidth]{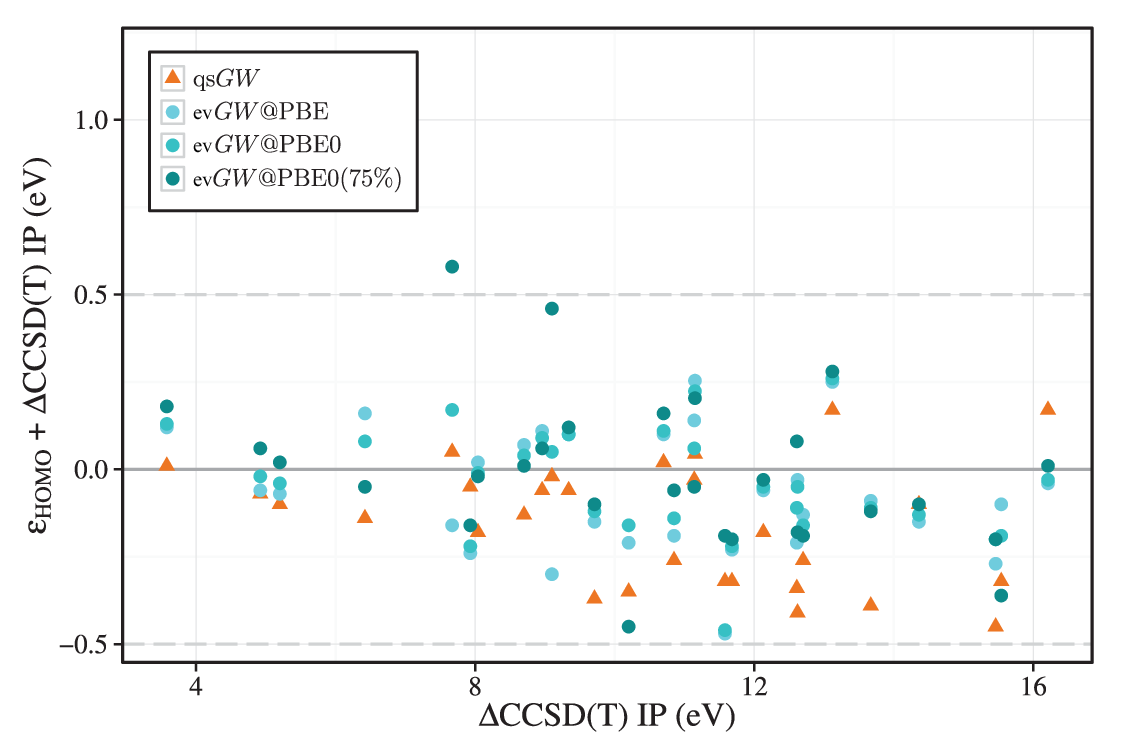} 
  \caption{Comparison of the  HOMO energies using ev$GW$ initialized from different DFT(PBE$\lambda$) starting points and qs$GW$ calculations to the $\Delta$CCSD(T)\cite{krause15} IP energies.}
  \label{fig:distance_evgw_ccsd} 
\end{figure}

To further substantiate the accuracy of ev$GW$ we present a comparison to $\Delta$CCSD(T)\cite{krause15} first IPs over the full test set, see \figref{fig:distance_evgw_ccsd}. The statistical analysis, see \tabref{tab:statisticsCCSD_IP_evgw}, shows that the overall error is not larger than the error from qs$GW$. In fact ev$GW$ results exhibit an even better agreement with $\Delta$CCSD(T), especially when starting from PBE0. Starting from PBE0(75\%), PBE0 with 75\% exact exchange,  introduces various spurious outliers. Summarizing, ev$GW$ seems to be a promising alternative to qs$GW$ leading in practice to results very close to the $\Delta$CCSD(T) reference, with a typical deviation of a few 100~meV.

\begin{table}[!htb]
\centering
\begin{tabular}{lrrrr}
\hline\hline
& & \multicolumn{3}{c}{ev$GW$}\\
& qsGW &  \quad @PBE & @PBE0  & @PBE0(75\%)\\
\hline
ME              &-0.16 &-0.07 &-0.04 &-0.04 \\ 
MAE            & 0.19 & 0.15 & 0.12 & 0.18 \\ 
$\sigma^2$ & 0.02 & 0.01 & 0.01 & 0.06 \\ 
MaxAE        & 0.45 & 0.47 & 0.46 & 1.17\\ 
MinAE         & 0.01 & 0.02 & 0.01 & 0.01\\ 
\hline\hline
\end{tabular} \caption[ev$GW$ first IPs vs. $\Delta$CCSD(T)]{Statistical evaluation of the data shown in \figref{fig:distance_evgw_ccsd}. The ev$GW$ approach shows better agreement with the $\Delta$CCSD(T) reference than the qs$GW$ approach, for all considered starting points. Best agreement of ev$GW$ with $\Delta$CCSD(T) is achieved by employing the PBE0 functional in the starting point for the DFT calculation. All values are in eV.}\label{tab:statisticsCCSD_IP_evgw}
\end{table} 

\subsection{Comparison of $G_{\text{ev}}W_0$ versus qs$GW$}\label{sect-gevw0}

So far, our results indicated that orbital updates have a minor effect on the  pole positions 
of qs$GW$. The next question is whether the QP-energies are as sensitive to the 
 pole positions of $W$ as they are to the pole positions  of $G$ in $\Sigma$. To this end we recall an approximate scheme, $G_{\text{ev}}W_0$, which we introduced before.\cite{kaplan15gwso} It keeps the screened interaction based on the starting point calculation $W_0$, and introduce self-consistency only at the  pole positions of the Green's function.  The computational cost of $G_{\text{ev}}W_0$ is the same as traditional $G_0W_0$ (within our implementation). 

\subsubsection{First Ionization Potentials}

\begin{figure}[!htb]
   \includegraphics[width=0.95\columnwidth]{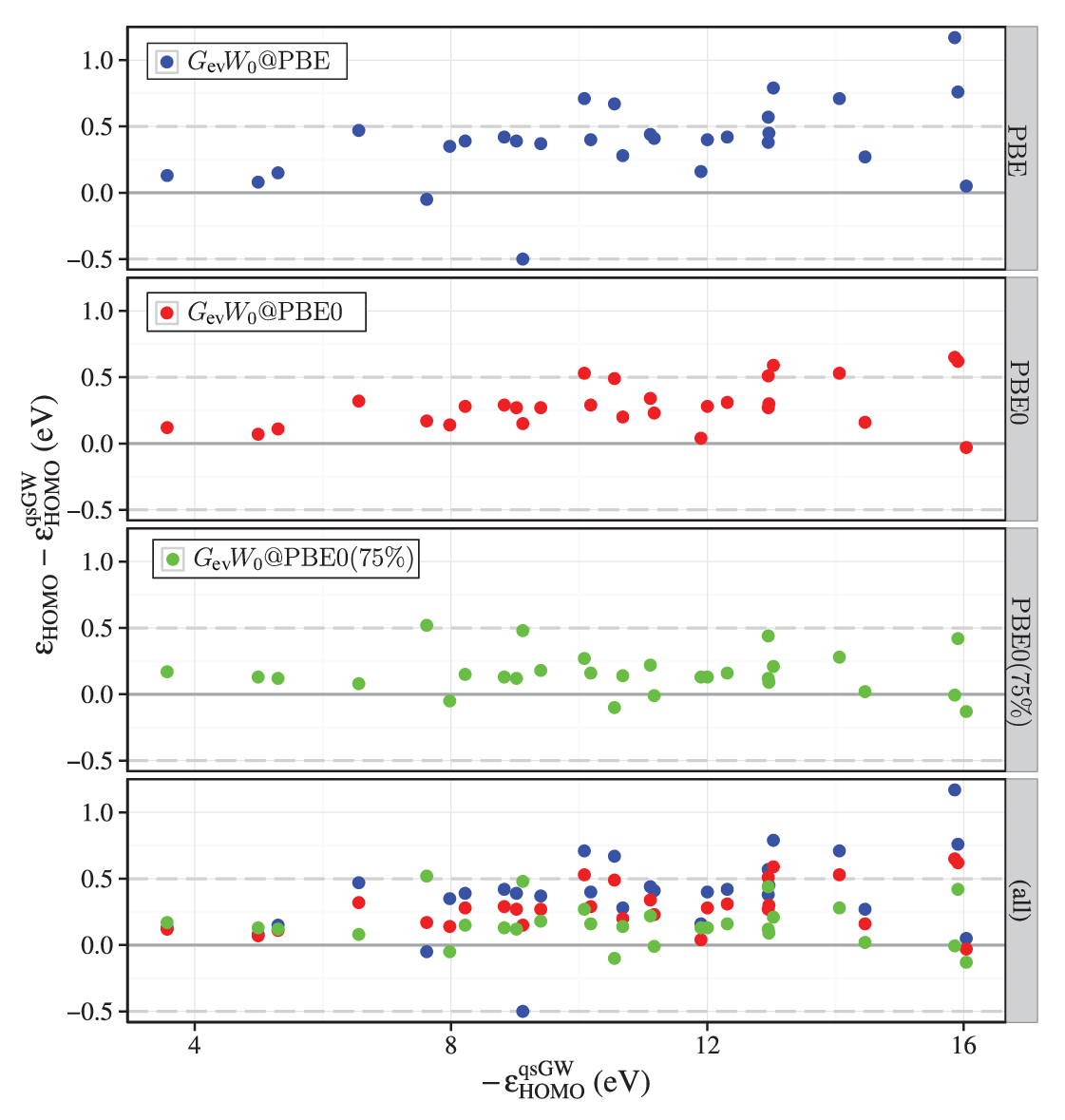} 
  \caption{Distance of calculated HOMO energies using  $G_{\text{ev}}W_0$ from DFT(PBE$\lambda$) starting points to the qs$GW$ HOMO energies. With increasing exact exchange contribution in the PBE$\lambda$ starting point improves the agreement with qs$GW$. Nevertheless, we find for all starting points an error which increase with energy. the statistical evaluation is reported in \tabref{tab:statistics_evGW0_qsgw}.}
  \label{fig:distance_evGW0_qsGW} 
\end{figure}

\figref{fig:distance_evGW0_qsGW} shows the difference of the HOMO energy obtained as from $G_{\text{ev}}W_0$ to the qs$GW$ HOMO energies for the full test set. The statistical measures are reported in \tabref{tab:statistics_evGW0_qsgw}. The first IPs from $G_{\text{ev}}W_0$ show a reduced starting point dependence in comparison to the single-shot $G_0W_0$ and the $G_0W_0$2nd results. However, the starting point dependence is much stronger that in the case of ev$GW$.  Furthermore, $G_{\text{ev}}W_0$ shows improved agreement with qs$GW$ if one employs the PBE or PBE0 starting point. If one employs the optimal PBE0(75\%) starting points the plain $G_0W_0$ and the $G_{\text{ev}}W_0$ are comparable close to the qs$GW$ results.

Comparing the optimal staring point we find for the three different approximate GW flavors, $G_0W_0$ ev$GW$ and $G_{\textrm{eV}}W_0$ there are some interesting remarks to make.  In ev$GW$ the differences are rather small with a small preference of PBE0, which is in agreement with PBE0 having the best dipole moment of the various starting points. The results are comparable to those of qs$GW$. The exact shape of the orbitals hence does not impact the results that much. However, if $W$ is kept fixed there is a clear preference for a PBE0(75\%) starting point. This indicates that given the, relatively small, spread in orbitals obtained in the range from PBE to HF it is primarily important to have a good screening. Good orbitals is only of secondary importance. This is also reflected in the small orbital corrections obtained in qs$GW$.

\begin{table}[!htb]
\centering
\begin{tabular}{lcccc}
\hline\hline
$G_{\text{ev}}W_0$ & @PBE &  @PBE0  &   @PBE0(75\%) \\
\hline
ME & 0.39 & 0.29 & 0.13 \\ 
MAE & 0.43 & 0.30 & 0.21 \\ 
$\sigma^2$ & 0.09 & 0.03 & 0.06 \\ 
MaxAE & 1.18 & 0.66 & 0.83 \\ 
MinAE & 0.04 & 0.04 & 0.01 \\ 
\hline\hline
\end{tabular} \caption[$G_{\text{ev}}W_0$ first IPs vs. qs$GW$]{Statistical measures of the difference of the calculated $G_{\text{ev}}W_0$ HOMO energies from DFT(PBE$\lambda$) starting points to the qs$GW$ HOMO cumulated over the test set. Best agreement is achieved employing the PBE0(75\%) starting point. All values are in eV.}\label{tab:statistics_evGW0_qsgw}
\end{table}

\subsubsection{Higher Ionization Potentials}

\begin{figure}[bth]
  \includegraphics[width=0.95\columnwidth]{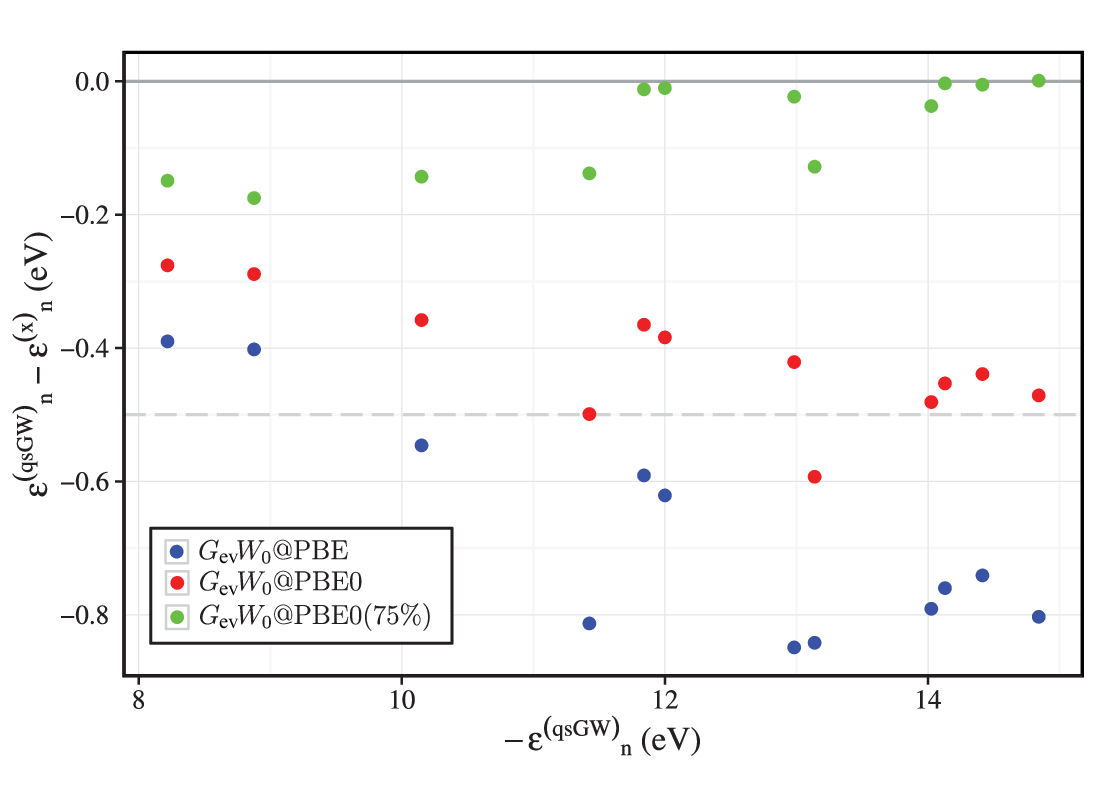}
   \caption{Deviation of the QP-energies from $G_{\text{ev}}W_0$ to QP-energies from qs$GW$ for naphthalene. Three different DFT based starting points employing the PBE, and the PBE hybrid functional with 25\% (PBE0) and with 75\% exact exchange(PBE0(75\%)) were chosen. The best agreement, with qs$GW,$ yields the calculation with the PBE0(75\%) starting point.} \label{fig:gevw0_naphthalene}
\end{figure}

In \figref{fig:gevw0_naphthalene} the QP-energies from $G_{\text{ev}}W_0$ are compared to the ones from qs$GW$ for naphthalene. The spectrum based on $G_{\text{ev}}W_0$ shows a slight overestimation of  the QP-energies for all starting points. Furthermore, we find that the shift on the poles from the self-energy due to the QP-correction corrects to poles towards the full qs$GW$ solution. Regarding the starting point dependence we find again that the PBE0(75\%) gives best agreement with the qs$GW$ results.

\section{Computational Performance}
\label{sect-compperf}

\begin{figure}[h]
   \includegraphics[width=0.95\columnwidth]{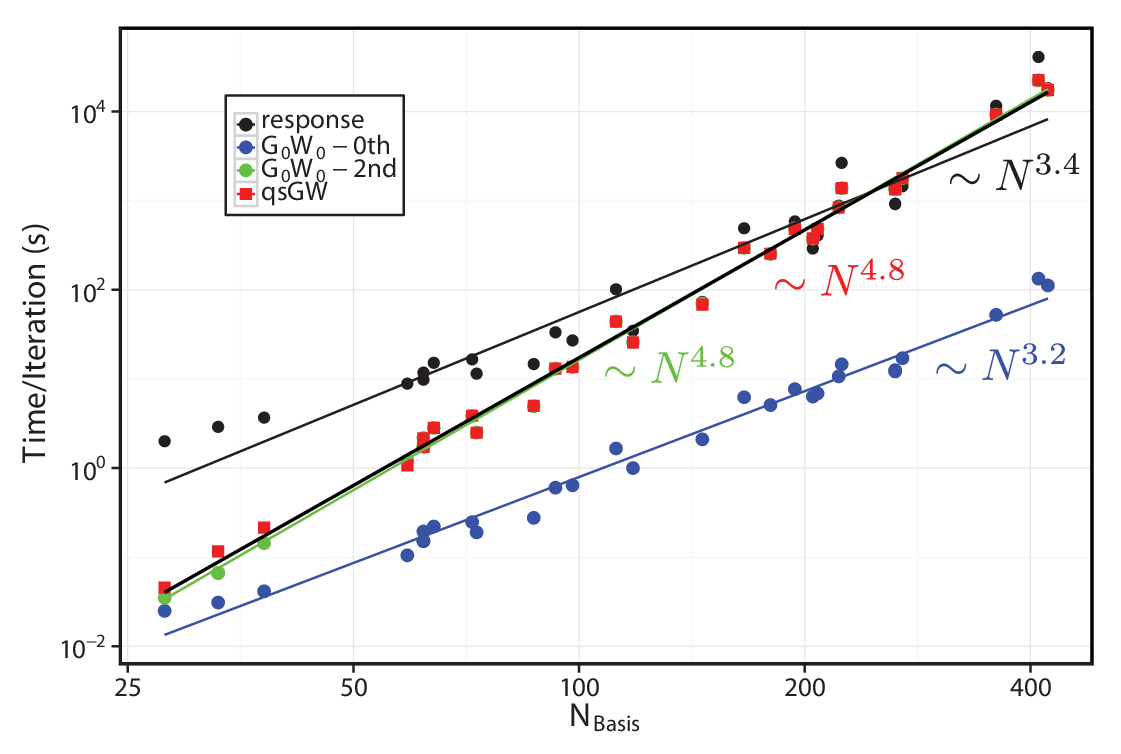} 
  \caption{The computational time needed for the construction of the response function and the time needed for a single iteration $GW$-proper over the number of basis functions. In ev$GW$ every full iteration consistes of a response step and a $G_0W_0$0th step. In $G_{\text{ev}}W_0$ only the first iteration has a response step, subsequent steps are computationally equivalent to a $G_0W_0$0th step, and times are almost the same for both procedures.}\label{fig:compTime} 
\end{figure}

The total computational effort needed for one $GW$ iteration comprises the calculation of the response function and the self-energy part. Under the latter we understand the construction of the self-energy and the solution of the quasi-particle equation. The time needed for the construction of the response function is for all considered approximations identical. The time needed for the self-energy part varies depending on the treatment of the QP-equation. \figref{fig:compTime} displays the computational cost for the different (approximate) $GW$ flavors within our implementation.

The $G_0W_0$0th approach is by far the fastest and shows the best scaling with the number of basis functions of only $\sim N^{3}$ for the self-energy part. The computational cost for $G_0W_0$0th calculations are hence dominated by the construction of the response function, which has a scaling of $\sim N^{4}$. However, this is only true for comparably small molecules. In the exact evaluation, one has to calculate all excitations, a matrix of range $\sim N^{2}$ has to be diagonalized. This is a hard $\sim N^{6}$-step, but with a small prefactor and still less than a $\sim N^{7}$ scaling for CCSD(T) in conventional implementations. 

Similar to $G_0W_0$, ev$GW$ approach operates solely on the diagonal part of the QP-equation. Hence, it has the same scaling behavior as (diagonal only) $G_0W_0$0th. But, to find a self-consistent solution, typically six iterations are necessary till convergence is achieved. Hence, the total time needed for ev$GW$ is that of $G_0W_0$ plus six times that of the the response calculation.

Both, $G_0W_0$2nd and the qs$GW$ approach, which diagonalize the QP-equation \ref{eq:QPE}, the self-energy part shows a scaling with the number of basis functions $N$ of $\sim N^{5}$, worse by a factor of $N^2$ as compared to $G_0W_0$0th and ev$G_0W_0$. Still we find that the computational time needed for the majority of the molecules within our test set is dominated by the construction of the response function. Only for systems requiring more than 340 basis functions, the self-energy part begins to dominate the computational cost. An actual optimization for best performance within the $GW$ is still an open task within our implementation. Especially in the qs$GW$ implementation we expect room for improvement.

Note that all $GW$-proper parts of our implementation have been designed for parallel computation. We find for all approaches a nearly perfect scaling with the number of cores, i.e. the typical setup of 8 core nodes gives a typical speedup of $7.94$. Hence, due to very good parallel scaling larger scale systems are tractable.

\section{Summary}
\label{sect-conclusions}

Considering the first IPs, we confirm for or a set of 29 representative molecules, that all flavors of the $GW$ method strongly improve agreement with experiment and CCSD(T) results as compared to the underlying DFT calculation. Moreover, the self-consistent qs$GW$ improves agreement with the CCSD(T) by an order of magnitude as compared to $G_0W_0$0th using a semi-local starting point (see \figref{fig:distanceHomo}). Also, we find that qs$GW$ has a better agreement with experiment in ionization spectra as compared to results based on $G_0W_0$  using a semi local starting point by typically 500~meV. Using an optional staring point, i.e. PBE0(75\%) for $G_0W_0$ and $G_\textrm{ev}W_0$ and PBE0 for ev$GW$, gives results very close to qs$GW$. Comparing qs$GW$ and sc$GW$, we observed a rigid shift towards stronger binding in qs$GW$, in the favor of qs$GW$ for both dipole moments and ionization energies.  The small number of molecules however prohibits to make a final claim un the ultimate  accuracy difference between sc and qs$GW$ for molecules. A study comparing the two using the $GW100$ set\cite{vansetten15gw100} is currently under way.

The qs$GW$ introduces only minor corrections on the spatial shape of the QP-orbitals and ground state density as compared to the KS-reference. Comparing calculated dipole moments to experimental results we find that the qs$GW$ improves agreement to experiment compared to traditional method like the Density Functional Theory (DFT) using a PBE functional, and the Hartree-Fock (HF) approach (see \tabref{tab:dipoles}). 

Based on this result we conclude that the partially self-consistent flavor of qs$GW$ updating only the orbitals, eigenvalue only qs$GW$ (ev$GW$) is an efficient alternative. The ev$GW$ is able to reproduce the (first) IPs from qs$GW$ very well and showed a strongly reduced dependence on the choice of the functional in comparison to $G_0W_0$. 

If system size is prohibiting even ev$GW$ we find the best alternative to be $G_0W_0$@PBE0(75\%). The increased amount of exact exchange is reasoned to mainly be necessary to improve the screening. Since the standard PBE0 orbitals turn out to be in general closest to the qs$GW$ results and also give best dipole moments as compared to experiment, one could even envision a hybrid approach. This would use the orbitals from PBE0, but perturbatively correct the eigenvalues using 75\% exact exchange before calculating the response function. 

\section{Outlook}

The full solution of Hedin's equations gives access not only to the quasi-particle (qp) energies, but also to charge neutral, particle-hole (ph) type excitations. A characteristic feature that all variants of the $GW$-theory share is that the energy of a ph-excitation is estimated neglecting the interaction between the particle and the hole, so that there is a tendency to overestimate the energy cost for creating ph-pairs, i.e. excitons.\cite{tddftanalog} As a result the Coulomb interaction is under screened, which leads to an underestimation of the magnitude of the correlation part of the self-energy. Thus, the HOMO energy is typically underestimated (i.e. the ionization potential is over estimated) as also observed in our results. 

There is no real consensus about the consequences of under screening for the qp-energies. A wide-spread belief is that under screening in the effective interaction $W$ (denominator) does not affect these energies because it is fully cancelled by the explicit vertex corrections  ($\Gamma$, nominator) appearing in the self-energy diagrams. Rigorous arguments in favor of the cancellation have not been given; so far, the existing evidence supporting cancellation relies on explicit calculations for a number of test systems, such as, e.g., homogeneous electron gases 
\cite{mahan89} and bulk silicon\cite{delsole94}.  

Our results indicate that when applying qs$GW$ to single molecules,  HOMO energies and ionization potentials are typically underestimated. In our opinion, a natural explanation is under screening of the Coulomb-interaction in $W$; we hence propose that the cancellation of vertex-terms is not efficient in molecular matter. 

To account for the binding-energy of excitons in the polarization function, one has to go beyond the $GW$-framework, i.e., include vertex corrections. A common means to this end is solving the Bethe-Salpeter equation (BSE). This is computationally challenging, however, and can be done only by employing suitable approximation schemes.\cite{sander15}

On the simplest level (RPA-X) one modifies the calculation of the polarization function by adding an exchange-matrix element to the Coulomb-matrix element of conventional RPA. The resulting theory would reproduce the time-dependent Hartree-Fock approximation if it were not for the correlation part of the self-energy. Following our philosophy of exploring the potential of static approximations, this step would be the most natural one to try next for us. 

Indeed, several groups have already employed the $GW$+BSE method to molecular systems in order to investigate into the effect of vertex corrections for charge-neutral excitations. We only mention two very recent works benchmarking the approach against TDDFT for small molecules.\cite{bruneval15,jacquemin15} We would like to consider these results as an indication that vertex-corrections are indeed significant for predicting and understanding molecular spectra.  These studies also indicate, that BSE-calculations inherit the starting point dependency of the underlying $\GoWo$-calculations.\cite{bruneval15} To eliminate this artifact of $\GoWo$, self-consistent calculations on the qs$GW$-level have been employed; they can achieve a numerical accuracy comparable to TD-PBE0 calculations.\cite{jacquemin15}
 
As promising as it is, the combination of $GW$ and BSE adds vertex corrections only to the response function still neglecting them when constructing $\Gamma$ for the self-energy $\Sigma$. The approximation thus obtained is not conserving. Moreover, if indeed large system classes exhibit a cancellation between vertex terms in $W$ and $\Gamma$, then these systems cannot be treated with the present technology. Therefore, we consider it very likely that a general method to be applied successfully to molecules and metals has to treat vertex corrections in $W$ and $\Sigma$ on the same footing.  

In principle, also the study of strongly correlated electron liquids relies on solving Hedin's equations -- at least to the extent that perturbative methods are being employed -- even though the respective community does not usually think about it in this way. In this context, a new method has been devised in recent years, the {\em functional renormalization group} method (\frg)\cite{SalmhoferBook,FrgReviewMetzner} Unlike $GW$+BSE, \frg includes vertex corrections to $W$ and $\Sigma$ in a consistent way. Its success for strongly-correlated matter originates from its capability to reliably signalize the existence of new phases with a broken symmetry, such as magnetism. From a conceptual point of view, \frg is very appealing because it can predict phase transitions in an unbiased manner;  in principle, an {\em apriori} guess about what phases are likely to appear is not required.

Similar to qs$GW$, also \frg is typically formulated employing a static approximation for the self-energy and the interaction vertex. Like traditional BSE-calculations, also the FRG focuses on the simplest subset of diagrams for the vertex.  However in contrast to traditional BSE, the \frg solves these (approximated) Hedin equations in a self-consistent manner. This is how FRG goes beyond the $GW$+BSE-scheme. 

The enhanced complexity requires a solution strategy that differs from the conventional iteration scheme to self-consistency still underlying, e.g., qs$GW$.  The strategy of \frg is to reformulate the self-consistency  problem in terms of a set of (non-linear) differential matrix equations.  The main idea may be understood as follows: In traditional  self-consistency solvers, an initial guess  gradually transforms (``flows'') into the fixed-point solution under the action of the iteration routine.   Now, \frg replaces this ``iterative flow'' by another flow along an artificial  coordinate, $\Lambda$, that plays the role of a cut-off energy familiar from  the renormalization group.  By construction, $\Lambda$ connects a known trivial solution of  Hedin's equations at $\Lambda{=}\infty$  with the exact solution at $\Lambda{=}0$.   The advantage of this formulation as compared to the traditional  self-consistency routines is that it is known exactly  how to initialize the flow equations; a starting guess for the Green's function, self-energy and interaction vertex is  not required. If instabilities occur along solving the  differential equation, then these can be interpreted as  ``runaway flow'' indicating a nearby phase with broken  symmetry. 

Applications of \frg to molecules do not yet exist. A first step into this direction has been made recently by two of us who formulated the \frg for systems without translational invariance.\cite{seiler16} Since vertex functions are kept explicitly, \frg comes with a computational complexity that might prohibit its use for intermediate sized molecular systems in the near future. Nevertheless, for benchmarking $GW$+BSE and high precision calculations for smaller molecules, \frg could have a  significant potential that we believe is worthwhile to explore in future research.

\begin{acknowledgement}
MvS acknowledges support financial support  the DFG-Center for Functional Nanostructures at the KIT. CS and FE acknowledge support from the DFG-grants EV30/7-1, EV30/8-1 and the DFG-Center for Functional Nanostructures at the KIT. FW and MEH also acknowledge the Bundesministerium fur Bildung und Forschung (BMBF) through the Helmholtz Research Program Science and Technology of Nanosystems for support and for providing the necessary infrastructure. Cpu time allocation at the HC3 cluster at the Karlsruhe Institute of Technology (KIT) Steinbuch Center for Computing (SCC) is gratefully acknowledged. The authors further acknowledge valuable discussions with X. Blase and V. Olevano. The authors also thank F. Caruso for providing the numerical data from ref.~\citenum{caruso12prb}.
\end{acknowledgement}

\begin{suppinfo}
Full details of the used atomic structures (TXT) and tables containing the calculated $\IP$'s and $\EA$'s as well as analysis of the computational performance and scaling scaling behavior of the competing methods (PDF). 
\end{suppinfo}

\bibliography{bibfile,notes}

\begin{mcitethebibliography}{106}
\bibitem[gw2()]{gw29}
The set of molecules used in this manuscript is based on the GW27 set exluding antracene and naphacene and adds Thiophene, Benzothiazole, Thiadiazole and Tetrathiofulvalene. These four are added to compare to existing full self-consistent GW results.\relax
\mciteBstWouldAddEndPunctfalse
\mciteSetBstMidEndSepPunct{\mcitedefaultmidpunct}
{}{\mcitedefaultseppunct}\relax
\EndOfBibitem
\bibitem[tdd()]{tddftanalog}
The situation is analogous in time dependent density functional theory when adiabatic approximations are being employed.\relax
\mciteBstWouldAddEndPunctfalse
\mciteSetBstMidEndSepPunct{\mcitedefaultmidpunct}
{}{\mcitedefaultseppunct}\relax
\EndOfBibitem
\end{mcitethebibliography}

\begin{figure}[h]
   \includegraphics[width=0.95\columnwidth]{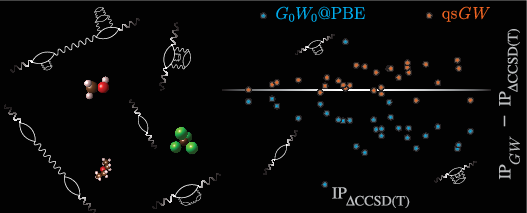} 
  \caption{TOC graphic}\label{TOCgraphic} 
\end{figure}

\end{document}